\documentclass[12pt]{article}
\usepackage{epsf}

\begin{document}

\newcommand{\gtrsim}{\mathop{}_{\textstyle \sim}^{\textstyle >}}
\newcommand{\lesssim}{\mathop{}_{\textstyle \sim}^{\textstyle <} }

\newcommand{\rem}[1]{{\bf #1}}

\begin{titlepage}

\def\thefootnote{\fnsymbol{footnote}}

\begin{center}

\hfill TU-812\\
\hfill April, 2008\\

\vskip .75in

{\Large \bf  
Big-Bang Nucleosynthesis and Gravitino }

\vskip .75in

{\large
$^{(a)(b)}$Masahiro Kawasaki, $^{(c)}$Kazunori Kohri, $^{(d)}$Takeo Moroi, 
\\
$^{(d)}$Akira Yotsuyanagi
}

\vskip 0.25in

{\em $^{(a)}$Institute for Cosmic Ray Research, University of Tokyo,
Kashiwa 277-8582, JAPAN}

\vskip 0.25in

{\em $^{(b)}$Institute for the Physics and Mathematics of the Universe,
University of Tokyo,\\
Kashiwa 277-8582, JAPAN}

\vskip 0.2in

{\em $^{(c)}$Physics Department, Lancaster University, Lancaster LA1 4YB,
UK}

\vskip 0.2in

{\em $^{(d)}$Department of Physics, Tohoku University,
Sendai 980-8578, JAPAN}

\end{center}
\vskip .5in

\begin{abstract}
    We derive big-bang nucleosynthesis (BBN) constraints on both
    unstable and stable gravitino taking account of recent progresses
    in theoretical study of the BBN processes as well as observations
    of primordial light-element abundances.  In the case of unstable
    gravitino, we set the upper limit on the reheating temperature
    assuming that the primordial gravitinos are mainly produced by the
    scattering processes of thermal particles.  For stable gravitino,
    we consider Bino, stau and sneutrino as the next-to-the-lightest
    supersymmetric particle and obtain constraints on their
    properties. Compared with the previous works, we improved the
    following points: (i) we use the most recent observational data,
    (ii) for gravitino production, we include contribution of the
    longitudinal component, and (iii) for the case with unstable
    long-lived stau, we estimate the bound-state effect of stau
    accurately by solving the Boltzmann equation.
\end{abstract}

\end{titlepage}

\renewcommand{\theequation}{\thesection.\arabic{equation}}
\renewcommand{\thepage}{\arabic{page}}
\setcounter{page}{1}
\renewcommand{\thefootnote}{\#\arabic{footnote}}
\setcounter{footnote}{0}

\section{Introduction}
\label{sec:intro}
\setcounter{equation}{0}

It is well known that, in the framework of the standard model of
particle physics, viable scenario of the evolution of the universe
cannot be obtained.  One of the most important reasons is that, in the
standard model, there is no candidate of dark matter.  Among various
possibilities, the lightest superparticle (LSP) in supersymmetric
models is a good candidate of dark matter, and hence supersymmetry is
thought to be very attractive from the cosmological point of view as
well as from particle-physics point of view.

In considering cosmology based on supersymmetric models with
$R$-parity conservation, it is important to be aware that graviton
also has its superpartner, which is called gravitino.  Even though
gravitino is an extremely weakly interacting particle, it may play a
significant role in the evolution of the universe. In particular, with
the existence of gravitino, there may exist a long-lived superparticle
whose lifetime is longer than $\sim 0.1\ {\rm sec}$.  For example, if
one of the superparticles of standard-model particles is the LSP,
gravitino becomes unstable and is very long-lived.  On the contrary,
if the gravitino is the LSP, lightest superparticle in the minimal
supersymmetric standard model (MSSM) sector, which we call MSSM-LSP,
decays into gravitino (and something else) with a Planck-scale
suppressed decay width.

Long-lived superparticles may affect the big-bang nucleosynthesis
(BBN) if their lifetimes are longer than $\sim 0.1\ {\rm sec}$.  Such
a long-lived superparticle is produced in the early universe and
decays in the thermal bath at the cosmic time comparable to its
lifetime.  Importantly, the energetic particles emitted by the decay
may induce hadronic and electro-magnetic shower and produce large
amount of energetic hadrons and/or photons.  Those energetic hadrons
and photons induce hadro- and photo-dissociation processes of light
elements which are synthesized by the standard BBN (SBBN) reactions.
In addition, mesons such as charged pions are also produced in the
hadronic shower induced by the decay of long-lived particle.  Such
charged pions and baryons cause extraordinary $p\leftrightarrow n$
conversion processes, resulting in the change of ${\rm ^4He}$
abundance.  Since the predictions of the SBBN scenario are in
reasonable agreements with observations, we obtain upper bound on the
primordial abundance of the long-lived superparticle.

In this paper, we study the BBN scenario in detail for the cases where
gravitino is unstable and stable, and derive constraints on those
cases.  In supersymmetric model, such BBN constraints have been
extensively studied for the cases where gravitino is unstable
\cite{Weinberg:1982zq,RadDecOld,RadDec,Kawasaki:1994sc,Kawasaki:2004yh,
Kawasaki:2004qu,Kohri:2005wn,EOV:2005} and stable
\cite{Moroi:1993mb,graLSP,Cyburt:2006,Jedamzik:2005dh,Kanzaki:2006hm}. The
primary purpose of this paper is to study the constraints on scenarios
with long-lived superparticles from the BBN scenario, taking account
of recent developments in this field.  Compared to the most recent
studies
\cite{Kawasaki:2004yh,Kawasaki:2004qu,Kohri:2005wn,Kanzaki:2006hm,
Kawasaki:2007xb}, we have improved the following points:
\begin{itemize}
\item We have used the most recent observational constraints on the
  primordial abundances of light elements.
\item We have taken into account the contribution of the production of
  the longitudinal component of gravitino in the calculation of the
  primordial abundance of gravitino.  Consequently, we have obtained
  severer upper bound on the reheating temperature for the case of
  unstable gravitino, in particular when the gravitino mass is
  relatively light.
\item For the case that the gravitino is the LSP and a charged slepton
  is the next-to-lightest supersymmetric particle (NLSP), we have
  calculated the abundance of the bound state of the charged slepton
  with $^4$He solving the Boltzmann equation without the use of Saha
  formula and hence have estimated the catalyzed production of $^6$Li
  precisely.  In addition, for the scenario with Bino-like NLSP, we
  have performed a detailed analysis.
\end{itemize}

The organization of this paper is as follows.  In Section
\ref{sec:bbn}, we first summarize how the BBN processes are analyzed
taking into account the effects of long-lived superparticles.  Then,
in Section \ref{sec:unstable}, we discuss constraints on the case
where gravitino is unstable.  We pay particular attention to the
scenario where the LSP is the lightest neutralino and derive the upper
bound on the reheating temperature after inflation.  In Section
\ref{sec:stable}, we consider the case where gravitino is the LSP and
hence is stable.  In such a case, the MSSM-LSP becomes unstable and
its decay may affect the BBN scenario.  The constraints depends on
what the MSSM-LSP is, so we consider several cases where the MSSM-LSP
is the lightest neutralino, one of the charged slepton, and one of the
sneutrino.  Section \ref{sec:conclusions} is devoted to conclusions
and discussion.

\section{BBN with Long-Lived Unstable Particles}
\label{sec:bbn}
\setcounter{equation}{0}

\subsection{Effects of the long-lived particle and calculation procedure}

We first summarize the general procedure to study effects of
long-lived unstable particles on BBN.  In particular, in this section,
we explain how we calculate non-standard BBN reaction rates with
long-lived unstable particles.  We also summarize observational
constraints on light-element abundances which will be adopted in our
study.

In our study, we consider two important types of decay modes of
long-lived particles: radiative decay and hadronic decay.  With the
radiative decay, photon and/or charged particles are emitted.  Then
the light elements would be photo-disintegrated (or photo-dissociated)
by energetic photons which are produced in the electro-magnetic shower
induced by high energy photons/charged
particles~\cite{RadDecOld,RadDec,Kawasaki:1994sc,EOV:2005,Kusakabe:2006hc}.
With the hadronic decay, the emitted hadrons induce extraordinary
interconversions between background proton and neutron ($p
\leftrightarrow n$ conversion) through the strong interaction.  Such
conversion processes enhance the neutron-to-proton ratio, resulting in
the over-production of
$^{4}$He~\cite{KaonAdp,Kawasaki:2004yh,Kawasaki:2004qu,Kohri:2005wn,
Jedamzik:2006xz}. The energetic nucleons also destroy the background
$^{4}$He and nonthermally produce other light elements, such as D, T,
$^{3}$He, $^{6}$Li, $^{7}$Li and $^{7}$Be~\cite{Dimopoulos:1987fz,
Dimopoulos:1988ue,Kawasaki:2004yh,Kawasaki:2004qu,Kohri:2005wn,
Jedamzik:2006xz}.  Furthermore, if the long-lived particle is charged,
it may form a bound state with background nuclei and change the
nuclear reaction rates.  In particular, for the production of ${\rm
^6Li}$, such a process, which is called catalyzed production process
\cite{Pospelov:2006sc}, may become very important.

We incorporate these effects, photo-dissociation, $p\leftrightarrow n$
conversion, hadro-dissociation, and catalyzed production process for
the case with charged long-lived particle, into a standard BBN code
\cite{Kawano:1992ua} and calculate light-element abundances.  (Some
details of these effects are discussed in the following.)  In order to
estimate errors in theoretical calculation, we perform Monte-Carlo
simulation by changing both the standard and non-standard reaction
rates within their statistical and systematic errors.  (For the
baryon-to-photon ratio, we use $\eta=6.10\pm 0.21$
\cite{Spergel:2006hy}.)  The estimated errors in the calculation of
light-element abundances are used in the estimation of the $\chi^2$
variable to derive BBN constraints; in the calculation of the $\chi^2$
variable, theoretical and observational errors are added in quadrature.

\subsubsection{Radiative decay}

In order to study the effects of radiative decay, we numerically solve
the Boltzmann equation governing the time evolution of the photon and
electron spectra. To obtain photo-dissociation rates, we take
convolutions of the photon spectrum and the energy-dependent cross
sections.  In the following, we briefly discuss our treatment of the
photo-dissociation processes.  (For the details of the effects of
radiative decay, see \cite{RadDec}.)

If the parent particle radiatively decays into high-energy photons or
charged particles, those daughter particles scatter off the background
photons $\gamma_{\rm BG}$ and electrons $e^{-}_{\rm BG}$ via
electromagnetic interaction.  (Here, the subscript ``BG'' is for
background particles in the thermal bath.)  The most important processes
are the Compton scattering ($\gamma + e^{-}_{\rm BG} \to \gamma +
e^{-}$), the electron-positron pair creation ($\gamma + \gamma_{\rm BG}
\to e^{-} + e^{+}$), the pair creation in nuclei ($\gamma + N \to e^{-}
+ e^{+} +N$, with $N$ being the background nuclei), the photon-photon
scattering ($\gamma + \gamma_{\rm BG} \to \gamma + \gamma$), and the
inverse Compton scattering ($e^{-} + \gamma_{\rm BG} \to e^{-} +
\gamma$).  Consequently, the electromagnetic cascade shower is induced
and a series of these processes produce many photons.

The timescale of the energy loss processes of the photons is much
faster than the timescale of cosmic expansion if the cosmic
temperature is sufficiently high $T>O(1\ {\rm eV})$.  On the other
hand, the thermalization timescale of the photons can be longer than
that to react with the background light elements, and hence the
photons may cause the photo-dissociation processes; the
photo-dissociation process with the threshold energy of $Q$ becomes
effective when the cosmic temperature is lower than $\sim
m_{e}^{2}/22Q$ \cite{Kawasaki:1994sc}.  For the photo-dissociation of
D ($^{4}$He), for which the threshold energy is $Q\sim 2.2 {\rm keV}$
(20 MeV), the cosmic temperature should be lower than $\sim 10 {\rm
keV}$ ($\sim 1 {\rm keV}$).

In addition, once the $^{4}$He is destructed, energetic T and $^{3}$He
are produced.  They scatter off background $^{4}$He and produce
$^{6}$Li nonthermally.  We also include photo-dissociation processes
of $^{6}$Li, $^{7}$Li and $^{7}$Be.

\subsubsection{Hadronic decay}

Next, we summarize how the effects of hadronic decay are studied.  If
the parent particle decays into quarks and/or gluons, they immediately
fragment into high-energy hadrons.  Comparing the lifetime of the
produced hadrons with the timescale of their interactions with the
background nuclei, only long-lived mesons, $\pi^{\pm}$, $K^{\pm}$,
$K^{0}_{L}$, as well as the nucleons $n$ and $p$ are important for the
study of the BBN reactions; hadrons with shorter lifetimes decay
before scattering off the background nuclei.

We calculate the distributions of partons emitted from the decaying
parent particle.  In particular, for the systematic study of the decay
chain, we use the ISAJET/ISASUSY packages \cite{Paige:2003mg}. Then
fragmentation processes of the partons into hadrons are simulated by
using the event generator PYTHIA \cite{Sjostrand:2000wi} and finally
the distributions of the produced hadrons are obtained.

The long-lived (and stable) hadrons destroy the background $^{4}$He
and produce D, T, and $^{3}$He copiously. Furthermore, the energetic
daughter T, $^{3}$He and the scattered $^{4}$He also nonthermally
produce $^{6}$Li, $^{7}$Li and $^{7}$Be through the collisions with
the background $^{4}$He. These processes are severely constrained by
the observations
\cite{Dimopoulos:1988ue,Kawasaki:2004yh,Kawasaki:2004qu}.

On the other hand, the neutron injection at around $T \sim 30$ keV may
reduce the final abundance of $^{7}$Be (or $^{7}$Li) through the
neutron capture process,
$^{7}$Be($n$,$p$)$^{7}$Li($p$,$\alpha$)$^{4}$He
\cite{Jedamzik:2004er,Kawasaki:2004qu,Kohri:2005wn,
Jedamzik:2005dh,Cumberbatch:2007me}.  However, this process does not
affect our constraints because we adopt a very mild observational
abundance of $^{7}$Li.

With the hadronic decay, the conversion processes of the background
$p$ and $n$ also occur.  In particular, once protons, neutrons, and
pions (and their anti-particles) are produced in the hadronic shower,
they induce $p \leftrightarrow n$ conversion processes and change the
neutron to proton ratio $n/p$ even after the normal freeze-out epoch
of neutron.\footnote
{We omit contributions from $K^{\pm}$ and $K^{0}_{L}$ to this
$p~\leftrightarrow n$ conversion according to a discussion in
\cite{Kawasaki:2004qu} although there have been some attempts to
include effects of Kaons \cite{KaonAdp,Kawasaki:2004yh}.}
Such $p\leftrightarrow n$ conversion processes increase ${\rm ^4He}$
and hence are constrained from observation of the primordial abundance
of ${\rm ^4He}$.

We solve the time-evolution of the hadron distribution functions
including all the relevant electromagnetic and hadronic energy-loss
processes, and calculate the reaction rates of the light elements
taking into account the hadro-dissociation and $p\leftrightarrow n$
conversion processes (as well as those of the photo-dissociation).
For further details, readers can refer to \cite{Kawasaki:2004qu}.

Before closing this subsection, we discuss the minor modification of
the treatment of the hadronic shower compared to the previous studies
\cite{Kawasaki:2004yh,Kawasaki:2004qu,Kohri:2005wn}.  In the present
study, we adopted a new evolution scheme of the hadronic shower.  For
the study of the abundance of non-thermally produced $^{6}$Li,
$^{7}$Li and $^{7}$Be, it is important to obtain information on the
energy distribution of $^{4}$He produced by the scattering process
$N+{\rm ^4He}\rightarrow N'+{\rm ^4He}+\pi$.  However, there is a lack
of experimental data of the energy distribution of the scattered
$^{4}$He in the inelastic $N+{\rm ^4He}$ collision.  In the old papers
\cite{Kawasaki:2004yh,Kawasaki:2004qu,Kohri:2005wn}, quantum
chromodynamics (QCD) properties was used to extrapolate results of
high-energy experiments to lower energies (as far as the extrapolation
is kinematically allowed).  Consequently, the final-state $^{4}$He
acquired larger energy than that estimated via equipartition
distribution in the center-of-mass system.  However, in the present
study, we adopted the smaller value of the $^4$He energy among the
equipartition value and the QCD prediction.  This procedure gives more
conservative constraints.  For non-relativistic $N+{\rm ^4He}$
collisions, this modified method gives smaller energies of the
scattered $^{4}$He, which reduces the resultant nonthermally produced
Li and Be by $\sim 50\ \%$ or so \cite{Cumberbatch:2007me}.  However,
this does not affect the constraints given in
\cite{Kawasaki:2004yh,Kawasaki:2004qu,Kohri:2005wn}. That is because
the nonthermally produced Li and Be are subdominant compared to the
SBBN contributions, and also because we adopt conservative
observational constraints.

\subsection{Observational light-element abundances}

Next, we show observational constants on primordial abundances of D,
$^{3}$He $^{4}$He, $^{6}$Li and $^{7}$Li, which we adopt in our study.
The errors are presented at 1 $\sigma$.  The subscript ``p'' and
``obs'' are for the primordial and observational values, respectively.

We adopt the following observational constraints on the deuterium
abundance
\begin{eqnarray}
    \label{eq:Dobs}
    {\rm (n_{\rm D}/n_{\rm H})}_{\rm p} = 
    (2.82 \pm 0.26) \times 10^{-5},
\end{eqnarray}
which is the most-recently reported value of the weighted
mean \cite{O'Meara:2006mj}. This value well agrees with the baryon to
photon ratio suggested by the WMAP 3-year CMB anisotropy 
observation~\cite{Spergel:2006hy}. 

To constrain the primordial $^{3}$He abundance, we use an
observational $^{3}$He to D ratio as an upper bound, which is a
monotonically increasing function of the cosmic time.  (For details,
see \cite{Sigl:1995kk,Kawasaki:2004yh,Kawasaki:2004qu}.)  In this
study we adopt the newest values of D and $^{3}$He abundances
simultaneously observed in protosolar clouds (PSCs), $(n_{\rm
  ^3He}/n_{\rm H})_{\rm PSC} = (1.66 \pm 0.06) \times 10^{-5}$ and
$(n_{\rm D}/n_{\rm H})_{\rm PSC} = (2.00 \pm 0.35) \times
10^{-5}$~\cite{GG03}.  Then we get
\begin{eqnarray}
    \label{eq:He3D}
    (n_{\rm ^3He}/n_{\rm D})_{\rm p} < 0.83+0.27.
\end{eqnarray}

Concerning the mass fraction of $^{4}$He, in \cite{Izotov:2007ed}, two
values are recently reported by using old and new $^{4}$He-emissivity
data, $Y_{\rm p}$ = $0.2472 \pm 0.0012$ and $Y_{\rm p}$ = $0.2516 \pm
0.0011$, respectively.  It should be noted that the latter value is
inconsistent with the SBBN prediction ($\simeq 0.2484$) even 
adopting possible theoretical errors of $0.0004$.  However, we do not
take its face value since the error presented in \cite{Izotov:2007ed}
does not include systematic effects.  Thus, in this study, we add an
error of $0.0040$ \cite{Fukugita:2006xy} to derive conservative
constraint:
\begin{eqnarray}
    \label{eq:He4obs}
    Y_{\rm p} = 0.2516 \pm 0.0040.
\end{eqnarray}
(For the systematic uncertainty of the observed $^{4}$He abundance,
see also \cite{Peimbert:2007vm}.)

For $^{7}$Li, we adopt the most recent value of the $^{7}$Li to
hydrogen ratio $\log_{10}(^{7}$Li/H)$_{\rm obs}=-9.90\pm 0.09$ given
in \cite{Bonifacio:2006au}.  This is close to the value given in
\cite{RBOFN}: Log$_{10}(^{7}$Li/H)$_{\rm obs}=-9.91\pm 0.10$, and in
addition, slightly larger values have been also reported for years;
for example, see \cite{Melendez:2004ni}: Log$_{10}(^{7}$Li/H)$_{\rm
  obs}=-9.63\pm 0.06$.  Those observational values are smaller than
the SBBN prediction by approximately 0.3 dex or so.  Concerning the
inconsistency between the face value of the primordial $^{7}$Li
abundance and the SBBN prediction, various solutions have been
discussed from the viewpoint of both astrophysics
\cite{Pinsonneault:2001ub,Li7prod_Astro} and cosmology
\cite{Li7prod_Cos,Jedamzik:2004er,Kohri:2005wn,Jedamzik:2005dh,Jedamzik:2006xz,
Kohri:2006cn,Cyburt:2006,Bird:2007ge,Jittoh:2007fr,Cumberbatch:2007me,
Jedamzik:2007cp,Jedamzik:2007qk,Kusakabe:2007}.
However, the main purpose of this paper is to derive a conservative
constraint, so we do not go into the details of these models.
Instead, we add an additional systematic error of +0.3 dex into the
observational face value $n_{^{7}{\rm Li}}/n_{\rm H}$ in our study:
\begin{eqnarray}
    \label{eq:Li7obs}
    {\rm Log}_{10}(n_{^{7}{\rm Li}}/n_{\rm H})_{p}=
    -9.90 \pm 0.09 + 0.3.
\end{eqnarray}
Notice that $+0.3$~dex is the systematic error expected from effect of
$^7$Li depletion by rotational mixing in
stars \cite{Pinsonneault:2001ub}.\footnote
{See also
\cite{Yao:2006px,Fields:2006ga} for detailed discussion on possible
systematic uncertainties.}

As for a $^{6}$Li constraint, we use $(n_{\rm ^6Li}/n_{\rm ^{7}Li})_{\rm
obs}$ = 0.046 $\pm$ 0.022, which was newly-observed in a very metal-poor
star \cite{Asplund:2005yt}. We also add a systematic error of +0.084
\cite{Kanzaki:2006hm} to take into account depletion effects in stars
adopting the relation between $^7$Li and $^6$Li depletion factors,
$\Delta \log_{10}(n_{^{6}{\rm Li}}/n_{\rm H}) = 2.5 \Delta
\log_{10}(n_{^{7}{\rm Li}}/n_{\rm
H})$~\cite{Pinsonneault:1998nf,Pinsonneault:2001ub}, which leads to
$\Delta \log_{10}(n_{^{6}{\rm Li}}/n_{^{7}{\rm Li}}) = 0.45$ with $^7$Li
depletion factor $0.3$~dex.  Then we get the following upper bound:
\begin{eqnarray}
    \label{eq:Li6obs}
    \left( n_{\rm ^6Li}/n_{\rm ^{7}Li} \right)_p
     < 0.046 \pm 0.022 + 0.084.
\end{eqnarray}

\section{Unstable Gravitino}
\label{sec:unstable}
\setcounter{equation}{0}

\subsection{Primordial abundance of gravitino}

When gravitino is unstable, gravitinos produced in the early
universe may affect the light-element abundances because the lifetime
of gravitino is extremely long.  Indeed, if gravitino is lighter than
$\sim O(10\ {\rm TeV})$, its lifetime becomes longer than $\sim 1\rm
{\rm sec}$ and, in such a case, most of the primordial gravitinos
decay after the BBN starts.  Since the predictions of the SBBN are in
reasonable agreements with observations, the abundance of gravitino is
required to be small enough.  Thus, we obtain upper bound on the
abundance of primordial gravitino.

There are several possible mechanisms of producing gravitino in the
early universe; gravitino may be from the decay of condensations of
scalar fields, like moduli fields \cite{GravFromModuli} or inflaton
field \cite{GravFromInflaton}, as well as from scattering processes of
particles in the thermal bath.  Gravitino production due to the decay
of scalar condensations is, however, highly model-dependent.  In order
to derive conservative constraints, we presume that the primordial
gravitinos originate from the scattering processes.  In such a case,
the abundance of gravitino is approximately proportional to the
reheating temperature after inflation.  Then, in order not to spoil
the success of BBN scenario, upper bound on the reheating temperature
is derived.

Concentrating on the gravitino production by the scattering, the
evolution of the number density of gravitino is governed by the
following Boltzmann equation:
\begin{eqnarray}
  \frac{d n_{3/2}}{dt} =  - 3 H n_{3/2} + C_{3/2},
  \label{ngravdot}
\end{eqnarray}
where $H$ is the expansion rate of the universe and $C_{3/2}$ is the
collision term.  In our analysis, we study the above Boltzmann
equation simultaneously with other relevant equations:\footnote
{Here, we assume that the potential of the inflaton field is well
  approximated by parabolic potential.}
\begin{eqnarray}
  \frac{d \rho_{\rm inf}}{dt} &=& 
   - 3 H \rho_{\rm inf} - \Gamma_{\rm inf} \rho_{\rm inf},
  \\
  \frac{d \rho_{\rm rad}}{dt} &=& 
   - 4 H \rho_{\rm rad} + \Gamma_{\rm inf} \rho_{\rm inf},
   \label{rhoraddot}
\end{eqnarray}
where $\rho_{\rm inf}$ and $\rho_{\rm rad}$ are energy densities of
inflaton and radiation, respectively, and $\Gamma_{\rm inf}$ is the
decay rate of the inflaton field.  We numerically solve Eqs.\
(\ref{ngravdot}) $-$ (\ref{rhoraddot}) with initial conditions
$n_{3/2}=0$ and $\rho_{\rm inf}\gg\rho_X$; we follow the evolution of
$n_{3/2}$, $\rho_{\rm inf}$, and $\rho_{\rm rad}$ from the
inflaton-dominated epoch (i.e., $t\ll\Gamma_{\rm inf}^{-1}$) to the
radiation-dominated epoch (i.e., $t\gg\Gamma_{\rm inf}^{-1}$).  Most
of the energy density stored in the inflaton is converted to that of
radiation at the reheating temperature which is of the order of
$\sqrt{M_*\Gamma_{\rm inf}}$ (with $M_*\simeq 2.4\times 10^{18}\ {\rm
GeV}$ being the reduced Planck scale); in this paper, the reheating
temperature is defined as $3H(T_R) = \Gamma_{\rm inf}$ which gives 
\begin{eqnarray}
  T_{\rm R} \equiv
  \left(
  \frac{10}{g_*(T_{\rm R}) \pi^2} M_*^2 \Gamma_{\rm inf}^2
  \right)^{1/4},
  \label{T_R}
\end{eqnarray}
where $g_*(T_{\rm R})$ is the effective number of the massless degrees
of freedom at the time of reheating.  In our study, we use $g_*(T_{\rm
R})=228.75$.

The collision term has been calculated in
\cite{Bolz:2000fu,Pradler:2006hh} by taking account of the thermal
masses for gauge-boson propagators:
\begin{eqnarray}
  C_{3/2} = \frac{\zeta (3) T^6}{16\pi^3 M_*^2}
  \sum_{\rm a} c_a g_a^2  (T)
  \left( 1 + \frac{M_a^2 (T)}{3m_{3/2}^2} \right)
  \ln \frac{k_a}{g_a (T)},
  \label{CollisionTerm}
\end{eqnarray}
where the summation is over the standard-model gauge groups, and
$c_a=11$, $27$, and $72$, while $k_a=1.266$, $1.312$, and $1.271$, for
$U(1)_Y$, $SU(2)_L$, and $SU(3)_C$, respectively.\footnote
{See also \cite{Ferrantelli:2007bx} for possible corrections.}
In addition,
$M_a$ ($a=1-3$) denotes the gaugino mass. Compared to the previous
works on unstable gravitino \cite{Kawasaki:2004qu,Kohri:2005wn}, we
have included the production of the longitudinal mode, (part of) which
is from the term proportional to $M_a^2/m_{3/2}^2$ in Eq.\
(\ref{CollisionTerm}).  Although we numerically calculate the
primordial abundance of gravitino, we also provide its fitting
formula.  Let us define the yield variable for particle $X$ with
lifetime $\tau_X$ as
\begin{eqnarray}
  Y_X \equiv \left[ \frac{n_{X}}{s} \right]_{t\ll \tau_{X}},
\end{eqnarray}
where $n_X$ is the number density of $X$ and $s$ is the entropy
density.  Then, when the gaugino masses obey the grand-unified-theory
(GUT) relation, the yield variable of gravitino is well fitted by
\begin{eqnarray}
    \label{eq:Y32new}
  Y_{3/2} &\simeq&
  2.3 \times 10^{-14} \times T_{\rm R}^{(8)}
  \left[ 1 + 0.015 \ln T_{\rm R}^{(8)} - 0.0009 \ln^2 T_{\rm R}^{(8)}
    \right]
  \nonumber \\ &&
  + 1.5 \times 10^{-14} \times
  \left( \frac{m_{1/2}}{m_{3/2}} \right)^2  T_{\rm R}^{(8)}
  \left[ 1 - 0.037 \ln T_{\rm R}^{(8)} + 0.0009 \ln^2T_{\rm R}^{(8)}
    \right],
\end{eqnarray}
where $T_{\rm R}^{(8)}\equiv T_{\rm R}/10^8\ {\rm GeV}$ and $m_{1/2}$
is the unified gaugino mass at the GUT scale.  The above fitting
formula agrees well with the numerical result with the error of a few
\% for $10^{5}\ {\rm GeV}\leq T_{\rm R}\leq 10^{12}\ {\rm GeV}$.

\subsection{Models}

In order to study effects of photo- and hadro-dissociations and
$p\leftrightarrow n$ conversion, it is necessary to understand the
energy spectra of daughter particles produced by the decay of
gravitino.  If gravitino is not the LSP, gravitino decays into a
lighter superparticle and standard-model particle(s).  Those daughter
particles also decay if they are unstable.  The decay chain strongly
depends on the mass spectrum (and coupling constants) of the MSSM
particles.  Since there are so many parameters in the MSSM, it is
difficult to consider the whole parameter space.  Thus, we adopt four
parameter points of the so-called mSUGRA model to fix the MSSM
parameters.  Then, for those parameter points, we calculate decay
chain of the gravitino in details and derive BBN constraints.

In the mSUGRA model, all the MSSM parameters are determined by the
unified gaugino mass $m_{1/2}$, universal scalar mass $m_0$, universal
coefficient for the tri-linear scalar coupling $A_0$, ratio of the
vacuum expectation values of two Higgs bosons $\tan\beta$, and
supersymmetric Higgs mass $\mu_H$.  ($|\mu_H|$ is determined by the
condition of electro-weak symmetry breaking.)  The points we choose
are shown in Table \ref{table:mSUGRA}.  For the points we choose, we
also calculate the thermal relic density of the lightest neutralino,
which is the LSP; the resultant density parameters are also shown in
Table \ref{table:mSUGRA}.  (We use $h=0.73$ \cite{Spergel:2006hy},
where $h$ is the expansion rate of the universe in units of $100\ {\rm
  km/sec/Mpc}$.)  Notice that the Cases 1 and 2 are in the so-called
``co-annihilation region,'' the Case 3 is in the ``focus-point
region,'' and the Case 4 is in ``Higgs funnel region.''

\begin{table}[t]
  \begin{center}
    \begin{tabular}{lcccc}
      \hline\hline
      {} & {Case 1} & {Case 2} & {Case 3} & {Case 4}\\
      \hline
      $m_{1/2}$   
      & $300\ {\rm GeV}$  & $600\  {\rm GeV}$
      & $300\ {\rm GeV}$  & $1200\ {\rm GeV}$ \\
      $m_0$       
      & $141\  {\rm GeV}$ & $218\ {\rm GeV}$
      & $2397\ {\rm GeV}$ & $800\ {\rm GeV}$ \\
      $A_0$
      & $0$ & $0$ & $0$ & $0$ \\
      $\tan\beta$ & $30$ & $30$ & $30$ & $45$ \\
      $\mu_H$     
      & $389\ {\rm GeV}$ & $726\   {\rm GeV}$ 
      & $231\ {\rm GeV}$ & $-1315\ {\rm GeV}$ \\
      $m_{\chi_1^0}$
      & $117\ {\rm GeV}$ & $244\ {\rm GeV}$
      & $116\ {\rm GeV}$ & $509\ {\rm GeV}$ \\
      $\Omega_{\rm LSP}^{\rm (thermal)}h^2$ 
      & $0.111$ & $0.110$ & $0.106$ & $0.111$\\
      \hline\hline
    \end{tabular}
    \caption{mSUGRA parameters used in our analysis.}
    \label{table:mSUGRA}
  \end{center}
\end{table}

\begin{figure}[t]
  \begin{center}
    \centerline{{\vbox{\epsfxsize=0.5\textwidth\epsfbox{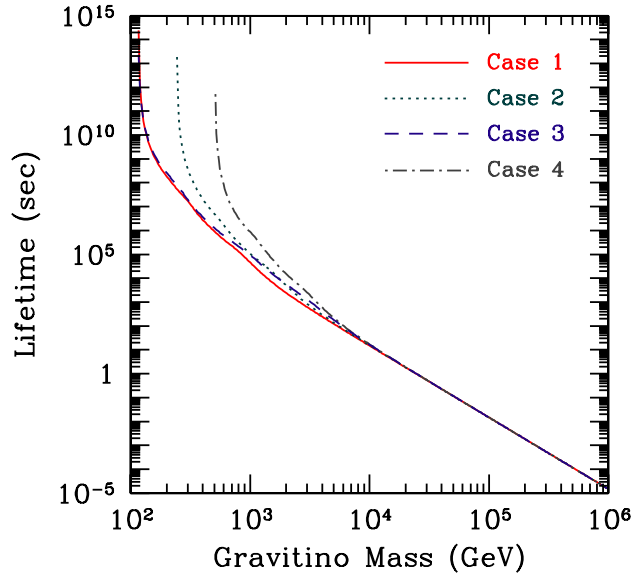}}}}
    \caption{Lifetime of gravitino for Cases 1 $-$ 4.  The
      horizontal axis is the gravitino mass.  }
    \label{fig:gravlife}
    \end{center}
\end{figure}

With the mSUGRA parameters given in Table \ref{table:mSUGRA}, the mass
spectrum of the MSSM particles as well as coupling parameters are
obtained.  Then, we calculate the decay rate of gravitino
$\Gamma_{3/2}$, taking account of all the relevant decay modes.  The
lifetime of gravitino is given by $\tau_{3/2}=\Gamma_{3/2}^{-1}$; we
show the lifetime for each cases in Fig.\ \ref{fig:gravlife}.  The
lifetime depends on the mass spectrum of the MSSM particles when the
gravitino mass is relatively light.  When gravitino becomes very
heavy, on the contrary, $\tau_{3/2}$ is determined only by the number
of possible decay modes.  In the present case, as one can see in Fig.\
\ref{fig:gravlife}, the lifetime becomes insensitive to the mass
spectrum of the final-state particles when $m_{3/2}\gtrsim 10\ {\rm
TeV}$.

The decay processes of the daughter particles from the gravitino decay
are also important; we also calculate partial decay rates of the
unstable superparticles to follow the decay chain.  The energy spectra
of the decay products are obtained by using Monte-Carlo analysis.  At
the parton level, the decay chain induced by the gravitino decay is
systematically followed by using ISAJET/ISASUSY
packages~\cite{Paige:2003mg}, while the hadronization processes of
partons are studied by using PYTHIA package~\cite{Sjostrand:2000wi}.
Then, we obtain energy spectra of (quasi) stable particles for the
study of the light-element abundances; once the spectra of hadronic
particles (in particular, those of proton, neutron, and pions) as well
as the averaged visible energy from the gravitino decay are obtained,
we calculate light-element abundances taking account of photo- and
hadro-dissociation processes as well as $p\leftrightarrow n$
conversion processes.

\subsection{Numerical results}

\begin{figure}
  \begin{center}
    \centerline{{\vbox{\epsfxsize=0.5\textwidth\epsfbox{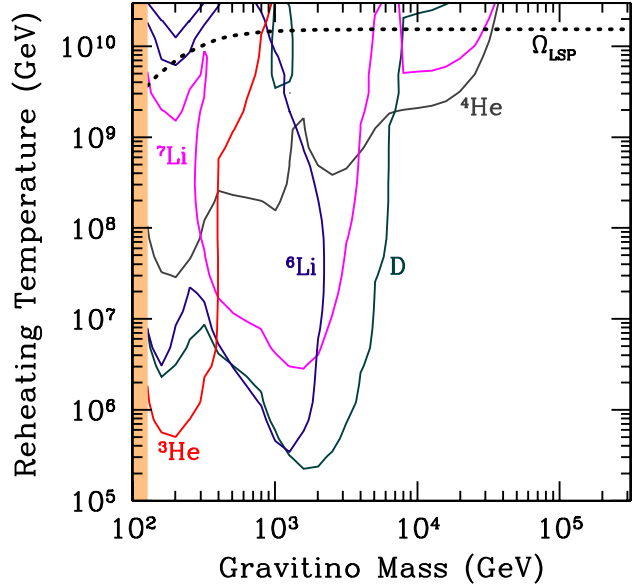}}}}
    \caption{BBN constraints for the Case 1 at 95\ \% C.L.  Each solid
      line shows upper bound on the reheating temperature from ${\rm
        D}$, ${\rm ^3He}$, ${\rm ^4He}$, ${\rm ^6Li}$, or ${\rm
        ^7Li}$.  The dotted line is the upper bound on the reheating
      temperature from the overclosure of the universe.}
    \label{fig:unstable01}
    \end{center}
\end{figure}

\begin{figure}
  \begin{center}
    \centerline{{\vbox{\epsfxsize=0.5\textwidth\epsfbox{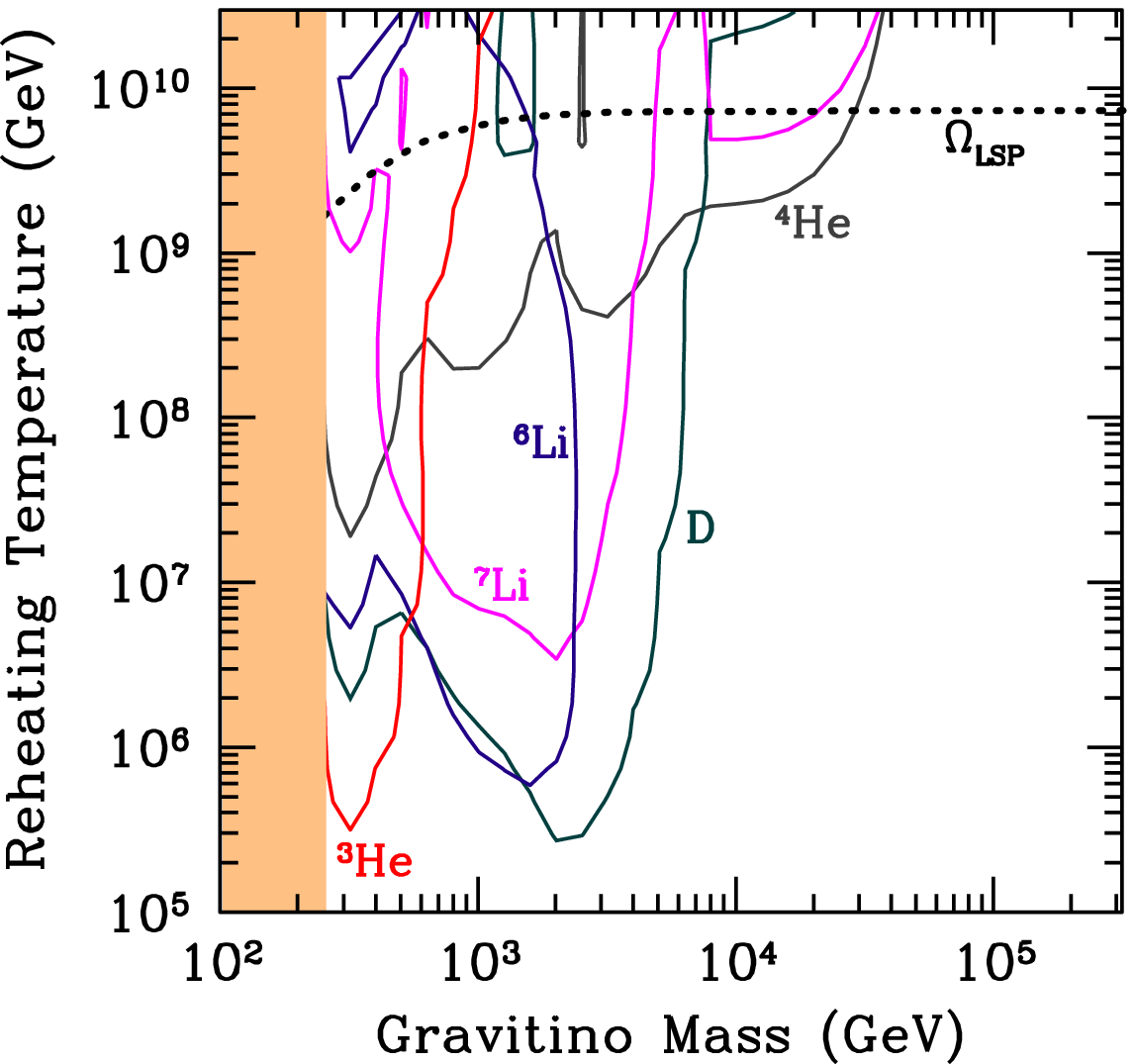}}}}
    \caption{BBN constraints for the Case 2.}
    \label{fig:unstable02}
    \end{center}
\end{figure}

\begin{figure}
  \begin{center}
    \centerline{{\vbox{\epsfxsize=0.5\textwidth\epsfbox{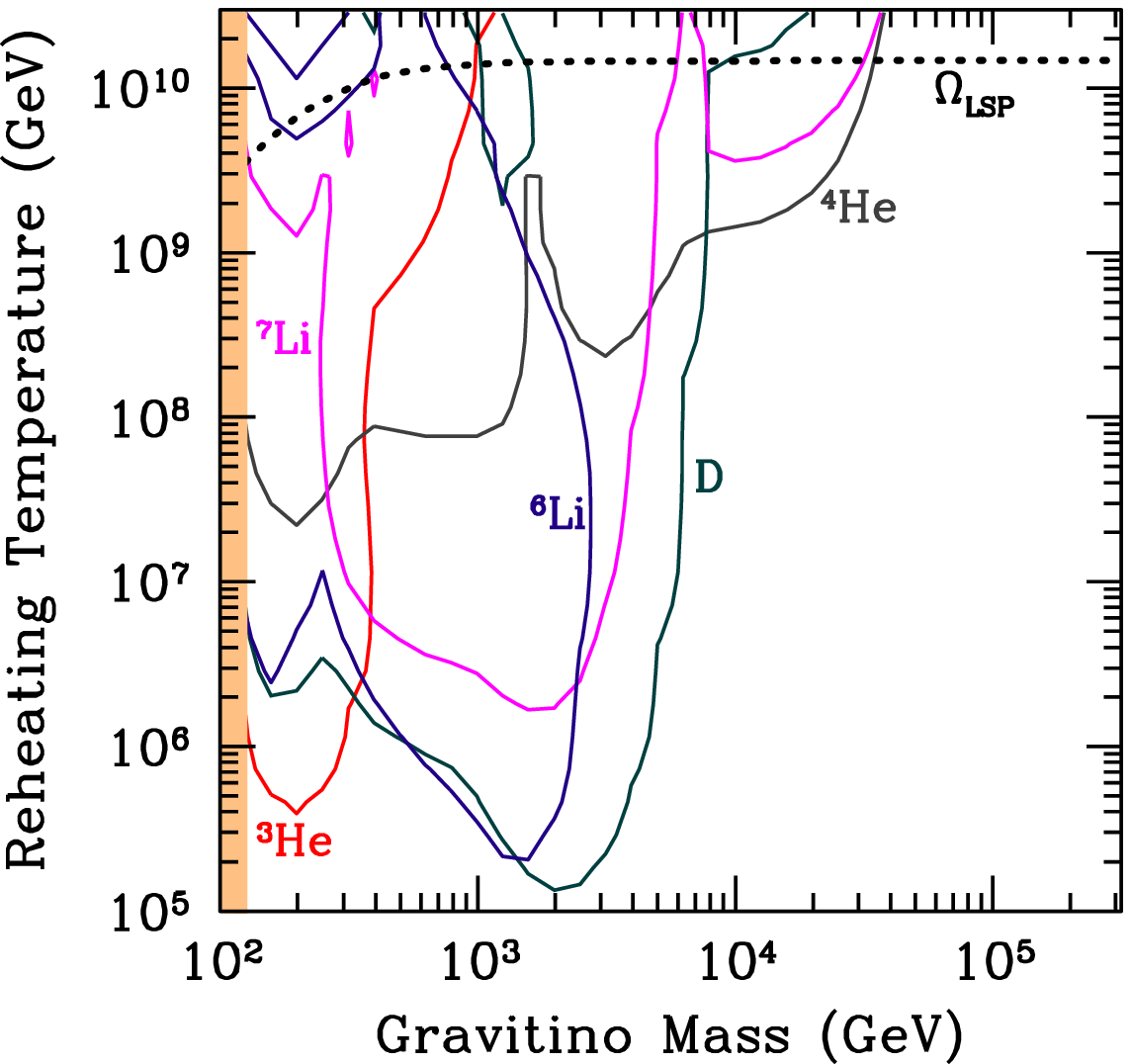}}}}
    \caption{BBN constraints for the Case 3.}
    \label{fig:unstable03}
    \end{center}
\end{figure}

\begin{figure}
  \begin{center}
    \centerline{{\vbox{\epsfxsize=0.5\textwidth\epsfbox{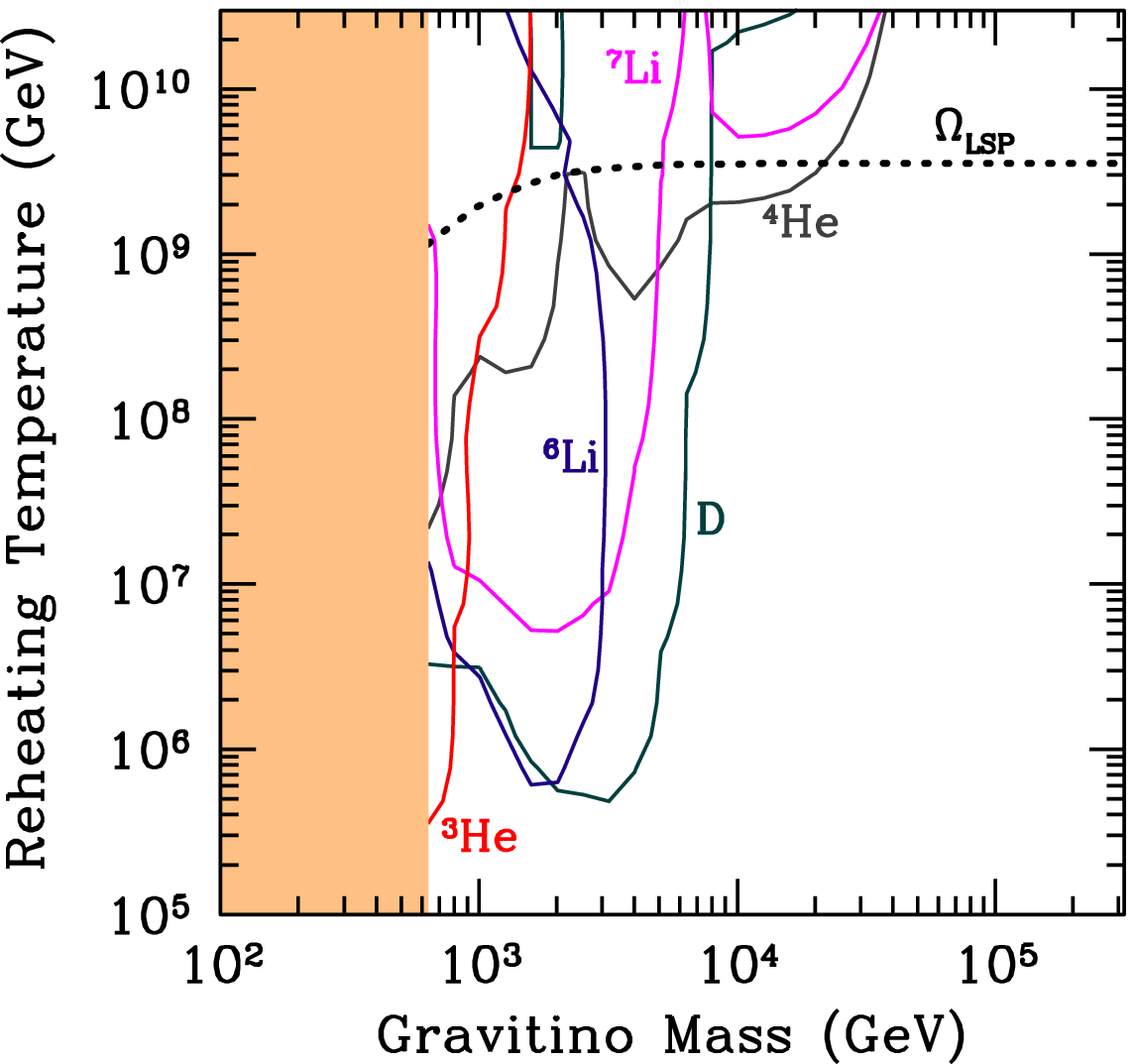}}}}
    \caption{BBN constraints for the Case 4.}
    \label{fig:unstable04}
    \end{center}
\end{figure}

We show our numerical results in Figs.\ \ref{fig:unstable01} $-$
\ref{fig:unstable04} for Cases 1 $-$ 4, respectively.  These figures
show upper bounds on the reheating temperature as functions of the
gravitino mass.  Since, in this section, we consider unstable
gravitino, we shade the region where the gravitino becomes lighter
than the MSSM-LSP and hence is stable.  Solid lines show upper bounds
on the reheating temperature from individual light elements; the
upper-left regions are excluded at 95 $\%$ C.L.  Compared with the
previous work~\cite{Kohri:2005wn}, we have newly included
contributions of the production of the longitudinal component of
gravitino.  (See Eq.\ (\ref{eq:Y32new})). This modification gives us
severer upper bounds on the reheating temperature when the gravitino
is relatively light.  Upper bound on the reheating temperature is
summarized in Table\ \ref{table:bound} for several values of the
gravitino mass .

\begin{table}[t]
  \begin{center}
    \begin{tabular}{l|llll}
      \hline\hline
      {$m_{3/2}$} & {Case 1} & {Case 2} & {Case 3} & {Case 4}\\
      \hline
      $300\ {\rm GeV}$
      & $1\times 10^{6}$ (${\rm ^3He}$)
      & $4\times 10^{5}$ (${\rm ^3He}$)
      & $1\times 10^{6}$ (${\rm ^3He}$)
      & $-$ 
      \\
      $1\ {\rm TeV}$
      & $5\times 10^{5}$ (${\rm ^6Li}$)
      & $9\times 10^{5}$ (${\rm ^6Li}$)
      & $3\times 10^{5}$ (${\rm ^6Li}$)
      & $3\times 10^{6}$ (${\rm ^6Li}$)
      \\
      $3\ {\rm TeV}$
      & $5\times 10^{5}$ (${\rm D}$)
      & $4\times 10^{5}$ (${\rm D}$)
      & $2\times 10^{5}$ (${\rm D}$)
      & $5\times 10^{5}$ (${\rm D}$)
      \\
      $10\ {\rm TeV}$
      & $2\times 10^{9}$ (${\rm ^4He}$)
      & $2\times 10^{9}$ (${\rm ^4He}$)
      & $2\times 10^{9}$ (${\rm ^4He}$)
      & $2\times 10^{9}$ (${\rm ^4He}$)
      \\
      $30\ {\rm TeV}$
      & $9\times 10^{9}$ (${\rm ^4He}$)
      & $8\times 10^{9}$ (${\rm ^4He}$)
      & $7\times 10^{9}$ (${\rm ^4He}$)
      & $8\times 10^{9}$ (${\rm ^4He}$)
      \\
      \hline\hline
    \end{tabular}
    \caption{Upper bound on the reheating temperature (in units of
      GeV) from BBN for Cases 1 $-$ 4.  The light element which gives
      the most stringent bound is indicated in the parenthesis.}
    \label{table:bound}
  \end{center}
\end{table}

When the gravitino is unstable, the LSP (i.e., the lightest neutralino
in this case) is produced by the decay of the gravitino.  Since the
decay of the gravitino occurs after the freeze-out epoch of the
lightest neutralino, the LSP produced by the gravitino decay survives
until today.  Thus, in this case, we also obtain the upper bound on
the reheating temperature from the overclosure of the universe.  The
lightest neutralino has two origins; one is thermal relic and the
other is non-thermal one from the gravitino decay.  However, since the
density of the thermal relic strongly depends on the MSSM parameters,
we do not take into account its effect in the calculation of the
density parameter.  Then, the density parameter of the LSP is
proportional to $Y_{3/2}$ and is given by
\begin{eqnarray}
  \Omega_{\rm LSP}h^2 \simeq 2.8\times 10^{10} \times Y_{3/2}
  \left(\frac{m_{\chi^0_1}}{100\ {\rm GeV}}\right),
\end{eqnarray}
where $m_{\chi^0_1}$ is the mass of the lightest neutralino (i.e., the
LSP) which is given in Table \ref{table:mSUGRA} for each cases.  We
require that $\Omega_{\rm LSP}h^2$ be smaller than the observed dark
matter density: $\Omega_{\rm LSP}h^{2} < 0.118$ (95 $\%$ C.L.)
\cite{Spergel:2006hy}, and derive upper bound on the reheating
temperature.  The bound is shown in the dotted line in the figures.
As one can see, the constraint from the relic density of the LSP is
less severe than those from BBN unless the gravitino mass is extremely
large.

In summary, when the gravitino mass is close the mass of the LSP, the
most stringent bound is from the overproduction of ${\rm ^3He}$,
resulting in the upper bound of $O(10^{5-6}\ {\rm GeV})$.  When
$m_{3/2}$ is around a few TeV, the hadro-dissociation processes become
most effective and the severest bound is from ${\rm D}$ or ${\rm
^6Li}$.  With higher gravitino mass, the constraint is drastically
relaxed since the gravitino lifetime becomes shorter.  In particular,
when the gravitino mass is larger than $\sim 30\ {\rm TeV}$ or so, the
reheating temperature can be as high as $\sim 10^{10}\ {\rm GeV}$.

\section{Stable Gravitino}
\label{sec:stable}
\setcounter{equation}{0}

\subsection{General remarks}

Next, we consider the case where gravitino is the LSP and hence is
stable.  In this case, the MSSM-LSP becomes unstable.  We assume that
the MSSM-LSP is the NLSP and that
it decays only into gravitino and standard-model particle(s).  Then,
the lifetime of the MSSM-LSP may become longer than $1\ {\rm sec}$.
If so, the decay products of the MSSM-LSP induce dissociation
processes and change the light-element abundances.  Since the lifetime
of the NLSP in this case is approximately proportional to $m_{3/2}^2$,
the scenario is constrained from BBN unless the gravitino mass is
small enough.  Importantly, the lifetime of the NLSP as well as the
spectra of charged particles (including hadrons) produced by the NLSP
decay are strongly dependent on what the NLSP is, so are the BBN
constraints.  Thus, in the following, we discuss several possible
cases separately.

\subsection{Case with neutralino NLSP}

First, we consider the case where the NLSP is the lightest neutralino
$\chi^0_1$.  Even in this case, in fact, properties of the NLSP
depends on MSSM parameters since the lightest neutralino is a linear
combination of Bino $\tilde{B}$, neutral Wino $\tilde{W}^0$, and
Higgsinos:
\begin{eqnarray}
  \chi^0_1 \equiv U_{\tilde{B}} \tilde{B} + U_{\tilde{W}} \tilde{W}^0
  + U_{\tilde{H}_u} \tilde{H}_u^0 + U_{\tilde{H}_d} \tilde{H}_d^0,
\end{eqnarray}
where $\tilde{H}_u^0$ and $\tilde{H}_d^0$ are up- and down-type
Higgsinos, respectively.  To make our discussion simple, we mostly
consider the case that the lightest neutralino is Bino-like (i.e.,
$U_{\tilde{B}}\simeq 1$ and $U_{\tilde{W}}\simeq U_{\tilde{H}_u}\simeq
U_{\tilde{H}_d}\simeq 0$).  Notice that the scenarios with Bino-like
lightest neutralino is realized easily.  In particular, if gaugino
masses obey the GUT relation, the Bino mass becomes lighter than Wino
and gluino masses.  In addition, assuming radiative electro-weak
symmetry breaking, the Higgsino mass $\mu_H$ becomes larger than $M_1$
in large fraction of the parameter space \cite{Drees:2004jm}.  With
those conditions, the lightest neutralino becomes Bino-like.

\begin{figure}[t]
  \begin{center}
    \centerline{{\vbox{\epsfxsize=0.55\textwidth\epsfbox{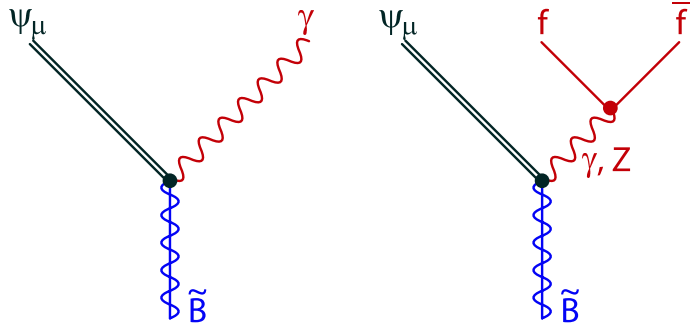}}}}
    \caption{Feynman diagrams relevant for the decay processes of
      Bino-like neutralino.}
    \label{fig:BinoDecay}
    \end{center}
\end{figure}

If $\chi^0_1\simeq\tilde{B}$, the relevant decay processes of the NLSP
is $\tilde{B}\rightarrow\psi_\mu\gamma$ and
$\tilde{B}\rightarrow\psi_\mu f\bar{f}$, where $\psi_\mu$ and $f$
denote gravitino and standard-model fermions, respectively.  (See
Fig.\ \ref{fig:BinoDecay}.)  Notice that the second process is
mediated by virtual and on-shell $Z$-boson as well as virtual photon
and hence, in our procedure, the decay process
$\tilde{B}\rightarrow\psi_\mu Z$ is taken into account in the process
$\tilde{B}\rightarrow\psi_\mu f\bar{f}$.  We calculate the decay rate
of these processes as well as the energy spectra of the final-state
particles, in particular, those of partons and electro-magnetic
particles.  Total decay rate of Bino-like neutralino, whose mass is
$m_{\tilde{B}}\simeq M_1$, is well approximated by
\begin{eqnarray}
  \tau_{\tilde{B}}^{-1} \simeq 
  \Gamma (\tilde{B}\rightarrow\psi_\mu\gamma) 
  + \Gamma (\tilde{B}\rightarrow\psi_\mu Z),
\end{eqnarray}
where \cite{Feng:2004mt}
\begin{eqnarray}
  \Gamma (\tilde{B}\rightarrow\psi_\mu\gamma) &=&
  \frac{\cos^2 \theta_{\rm W}}{48\pi M_*^2}
  \frac{m_{\tilde{B}}^5}{m_{3/2}^2}
  (1 - x_{3/2}^2)^3 (1 + 3 x_{3/2}^2),
  \\
  \Gamma (\tilde{B}\rightarrow\psi_\mu Z) &=&
  \frac{\sin^2 \theta_{\rm W} \beta_{\tilde{B}\rightarrow\psi_\mu Z}}
  {48\pi M_*^2}
  \frac{m_{\tilde{B}}^5}{m_{3/2}^2}
  \Big[
    (1 - x_{3/2}^2)^2 (1 + 3 x_{3/2}^2)
    \nonumber \\ &&
    - x_Z^2
    \left\{
      3 + x_{3/2}^3 (-12 + x_{3/2}) + x_Z^4 
      - x_Z^2 (3 - x_{3/2}^2)
    \right\} 
  \Big],
\end{eqnarray}
with $\theta_{\rm W}$ being the Weinberg angle, $x_{3/2}\equiv
m_{3/2}/m_{\tilde{B}}$, and $x_Z\equiv m_Z/m_{\tilde{B}}$.  In
addition, for $m_{\tilde{B}}>m_{3/2}+m_Z$,
\begin{eqnarray}
  \beta_{\tilde{B}\rightarrow\psi_\mu Z} \equiv
  \left[ 
    1 - 2 (x_{3/2}^2 + x_Z^2) 
    + (x_{3/2}^2 - x_Z^2)^2 
  \right]^{1/2},
\end{eqnarray}
while $\beta_{\tilde{B}\rightarrow\psi_\mu Z}=0$ otherwise.

As well as the decay rate and the energy spectra of decay products, it
is also necessary to obtain the primordial abundance of the lightest
neutralino in deriving the BBN constraints.  The primordial abundance
is determined at the time of its freeze-out if there is no entropy
production after freeze out, and is strongly dependent on MSSM
parameters.  In our study, instead of performing a precise calculation
of the primordial abundance assuming a detailed model, we first adopt
approximated formulas to derive constraints.  Following
\cite{Feng:2004mt}, for the abundance in the bulk region, we use
\begin{eqnarray}
  Y_{\tilde{B}} =
  4 \times 10^{-12} \times 
  \left( \frac{m_{\tilde{B}}}{100\ {\rm GeV}} \right) 
  \mbox{~~~: bulk},
  \label{Y_bino(bulk)}
\end{eqnarray}
while, for that in the focus-point and co-annihilation regions, we use
\begin{eqnarray}
  Y_{\tilde{B}} =
  9 \times 10^{-13} \times 
  \left( \frac{m_{\tilde{B}}}{100\ {\rm GeV}} \right)
  \mbox{~~~: focus / co-annihilation}.
  \label{Y_bino(focus)}
\end{eqnarray}

\begin{figure}
  \begin{center}
    \centerline{{\vbox{\epsfxsize=0.5\textwidth\epsfbox{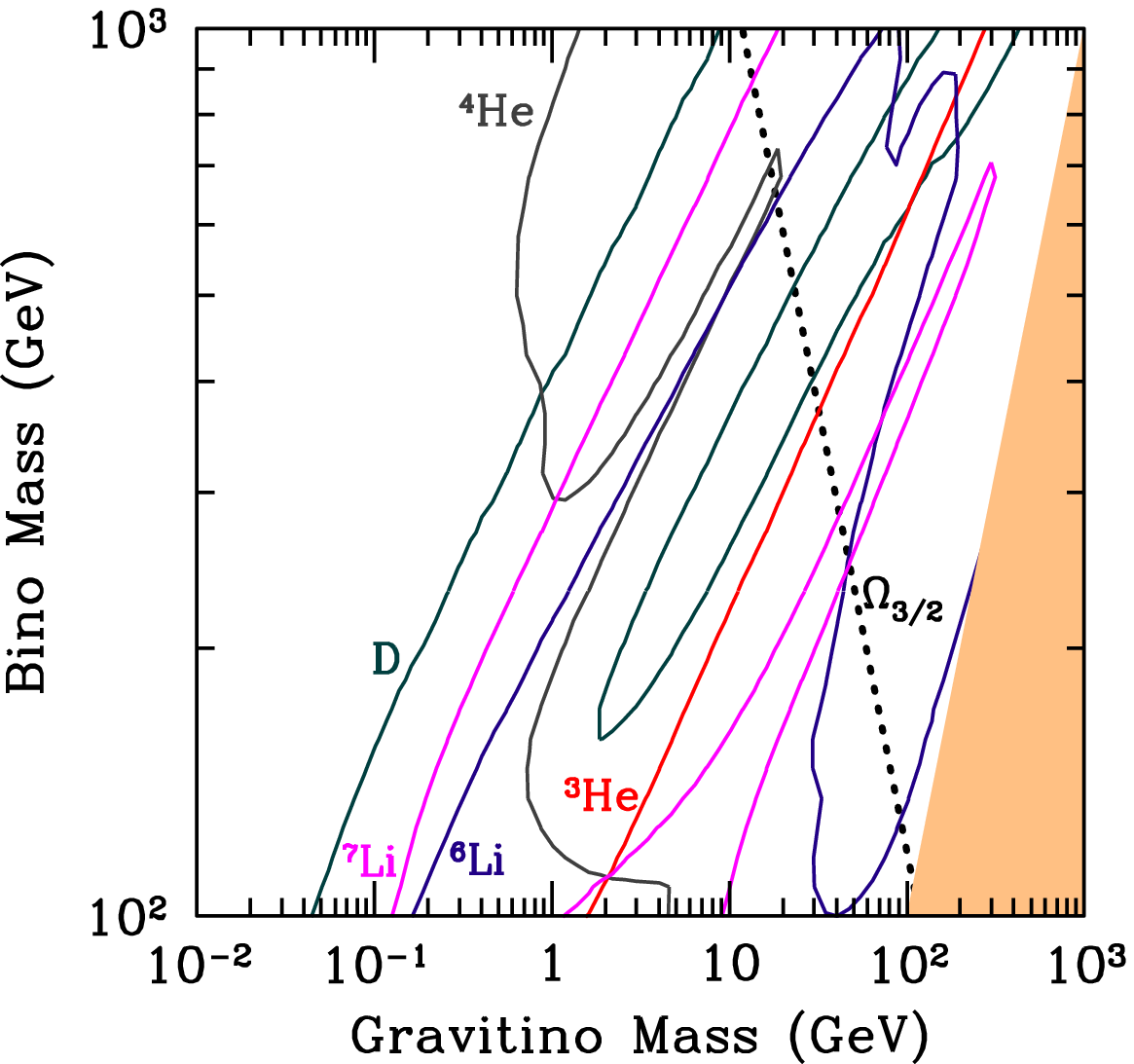}}}}
    \caption{BBN constraints for the Bino-NLSP scenario with the
      abundance given in Eq.\ (\ref{Y_bino(bulk)}).  In the shaded
      region, gravitino becomes heavier than the lightest neutralino
      and hence is not the LSP.  The region with small gravitino mass
      is allowed.}
    \label{fig:binobulk}
    \vspace{10mm}
    \centerline{{\vbox{\epsfxsize=0.5\textwidth\epsfbox{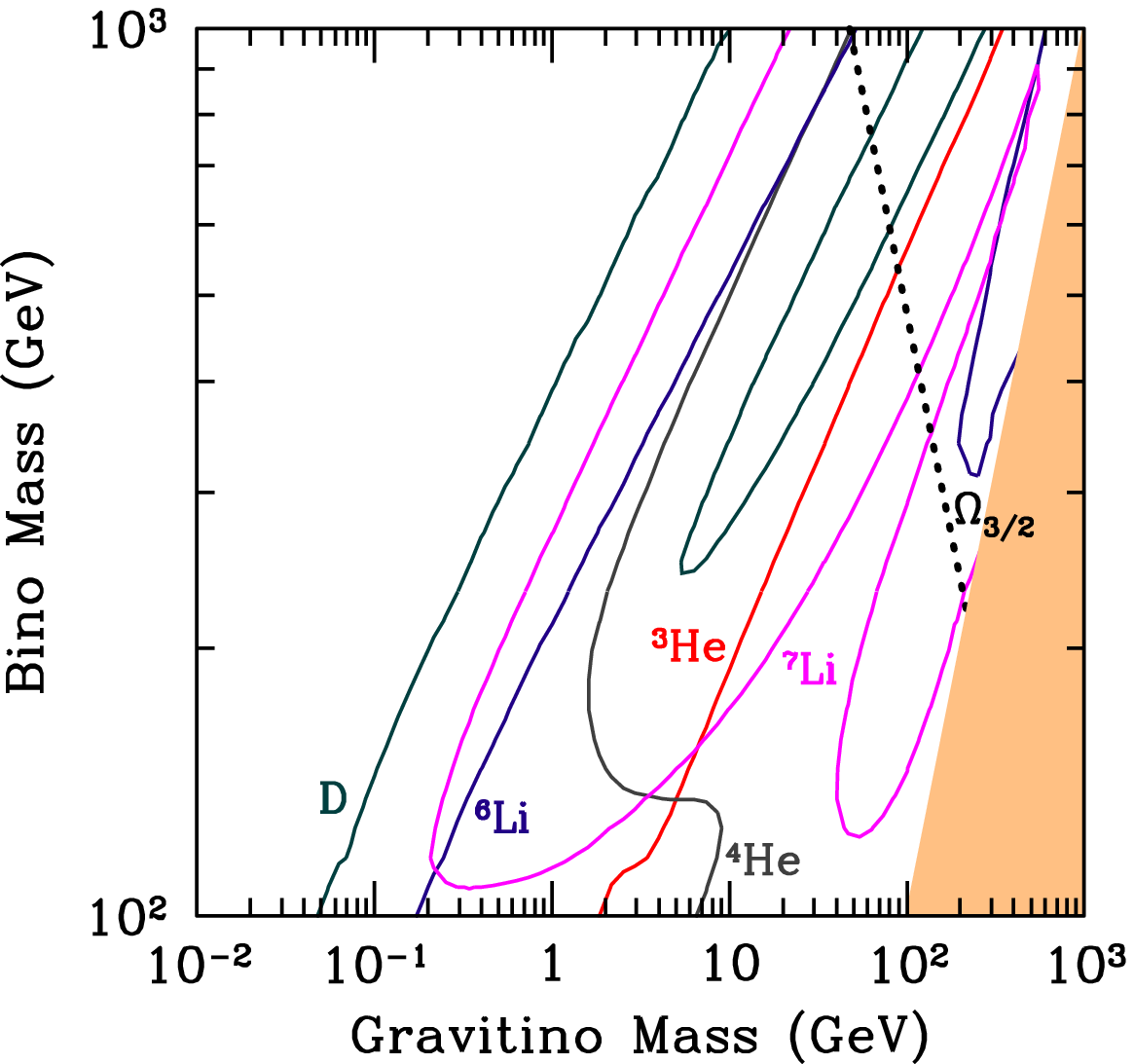}}}}
    \caption{Same as Fig.\ \ref{fig:binobulk}, except for the
      abundance given in Eq.\ (\ref{Y_bino(focus)}).}
    \label{fig:binofocus}
    \end{center}
\end{figure}

\begin{figure}
  \begin{center}
    \centerline{{\vbox{\epsfxsize=0.5\textwidth\epsfbox{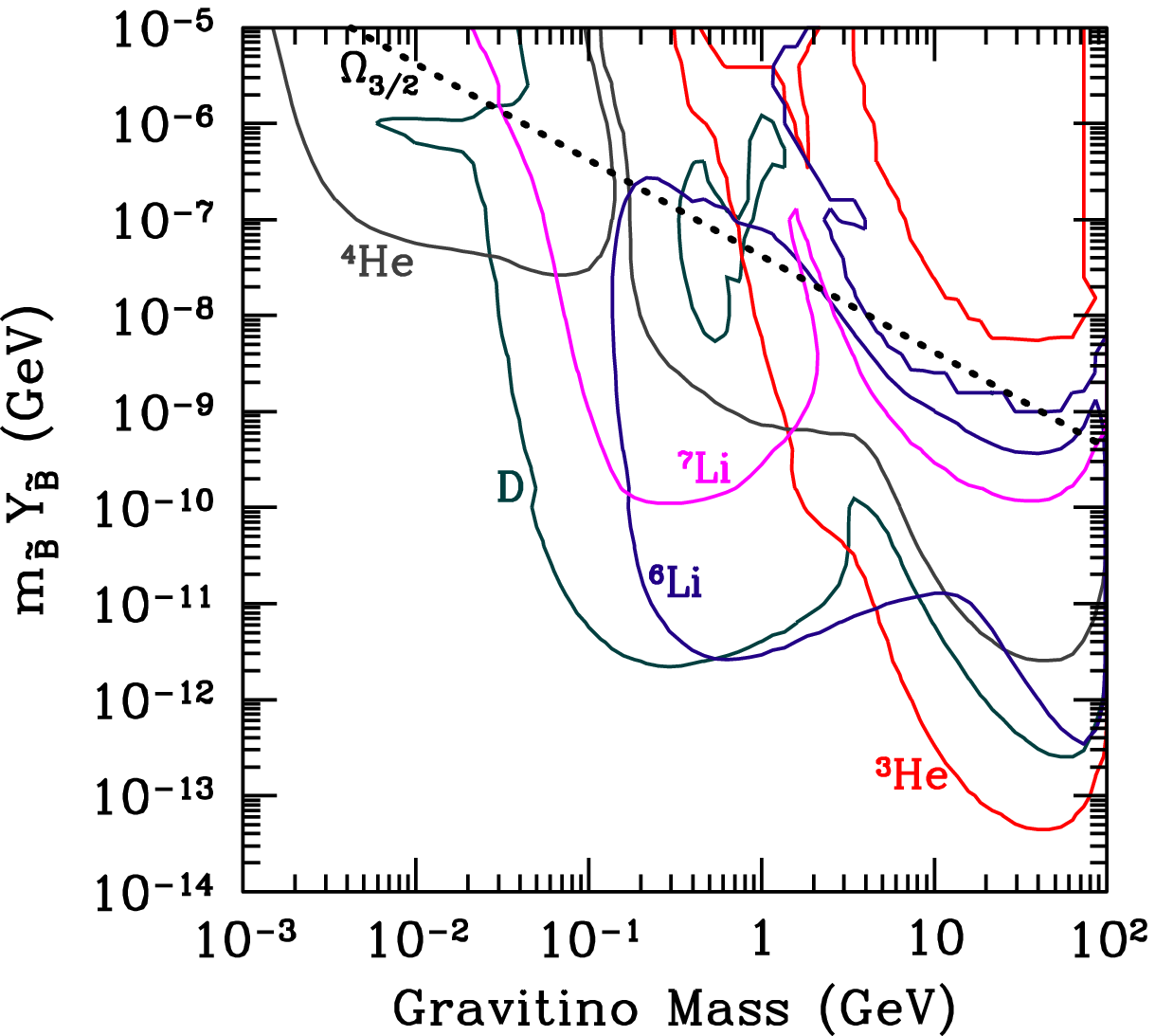}}}}
    \caption{BBN constraints for the Bino-NLSP scenario on the
      $m_{3/2}$ vs.\ $m_{\tilde{B}}Y_{\tilde{B}}$ plane.  The Bino
      mass is taken to be $100\ {\rm GeV}$.}
    \label{fig:binoM100}
    \vspace{10mm}
    \centerline{{\vbox{\epsfxsize=0.5\textwidth\epsfbox{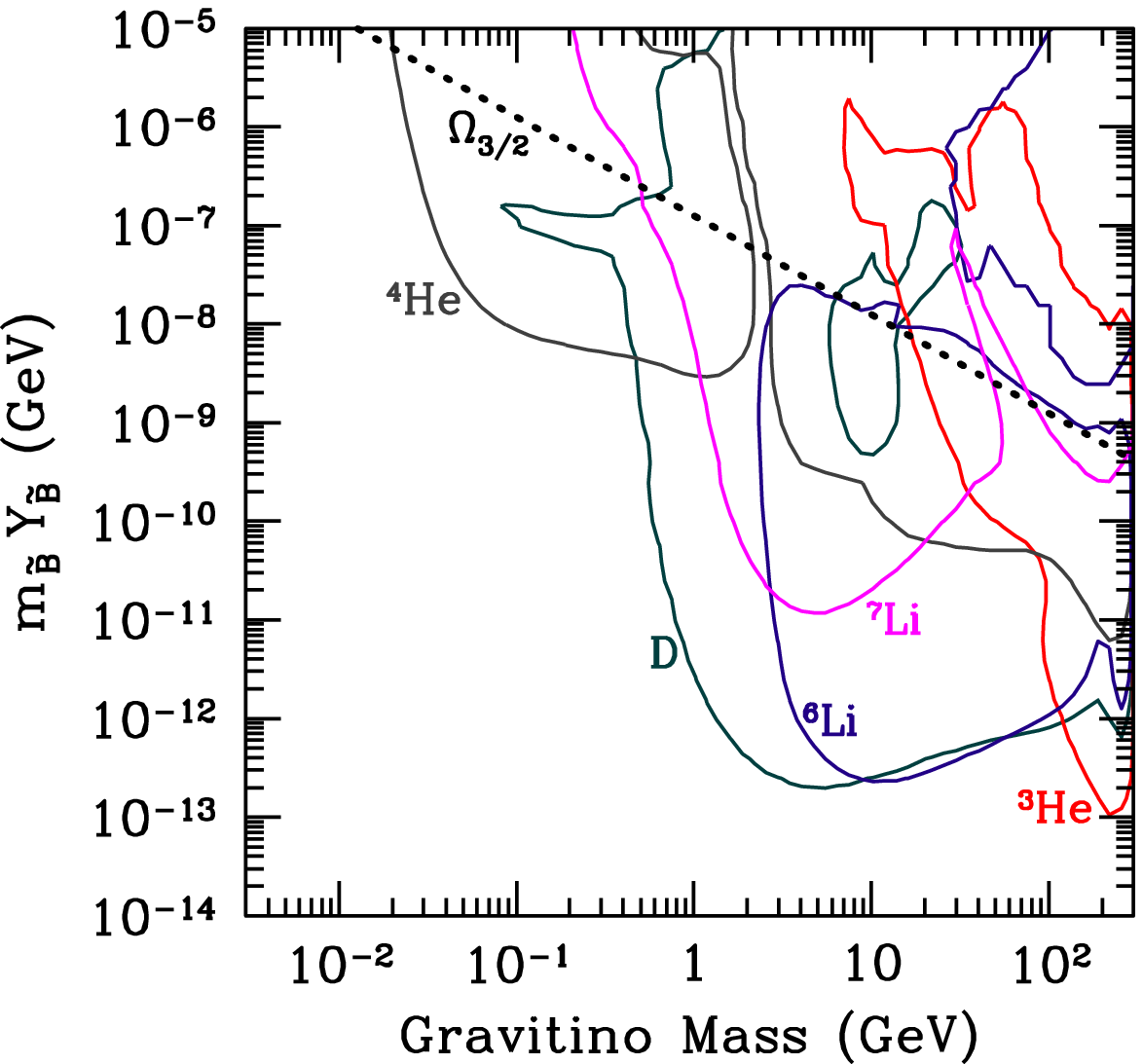}}}}
    \caption{Same as Fig.\ \ref{fig:binoM100}, except for 
      $m_{\tilde{B}}=300\ {\rm GeV}$.}
    \label{fig:binoM300}
    \end{center}
\end{figure}

Constraints on the $m_{3/2}$ vs.\ $m_{\tilde{B}}$ plane with the relic
abundance given in Eqs.\ (\ref{Y_bino(bulk)}) and
(\ref{Y_bino(focus)}) are shown in Figs.\ \ref{fig:binobulk} and
\ref{fig:binofocus}, respectively.  The most stringent constraint is
from the overproduction of ${\rm D}$ and ${\rm ^4He}$; overproduction
of ${\rm D}$ is due to the hadro-dissociation of ${\rm ^4He}$ produced
by the SBBN reactions, while that of ${\rm ^4He}$ is due to the
$p\leftrightarrow n$ conversion.  Comparing Figs.\ \ref{fig:binobulk}
and \ref{fig:binofocus}, we can see that the constraints from the
overproduction of ${\rm D}$ on the $m_{3/2}$ vs.\ $m_{\tilde{B}}$
plane are not so sensitive to $Y_{\tilde{B}}$.  This is because the
upper bound on the primordial abundance of $\tilde{B}$ from the
overproduction of ${\rm D}$ significantly changes with a sight change
of the lifetime.  (See Figs.\ \ref{fig:binoM100} and
\ref{fig:binoM300}.)  On the contrary, constraint from ${\rm ^4He}$ is
sensitive to $Y_{\tilde{B}}$; in the focus-point/co-annihilation case
where $Y_{\tilde{B}}$ is relatively small, constraint from ${\rm
^4He}$ is not important while, in the bulk case, ${\rm ^4He}$ is too
much produced when the Bino mass becomes large.

Next, we treat $Y_{\tilde{B}}$ as a free parameter and derive BBN
constraints on $m_{3/2}$ vs.\ $m_{\tilde{B}}Y_{\tilde{B}}$ plane.  The
results are shown in Figs.\ \ref{fig:binoM100} and \ref{fig:binoM300}
for $m_{\tilde{B}}=100\ {\rm GeV}$ and $m_{\tilde{B}}=300\ {\rm GeV}$,
respectively.  When the gravitino mass is close to $m_{\tilde{B}}$,
the lifetime of the lightest neutralino becomes relatively long.  In
such a case, the most stringent bound on $Y_{\tilde{B}}$ is from the
overproduction of ${\rm ^3He}$ due to the photo-dissociation of ${\rm
^4He}$.  On the contrary, with smaller gravitino mass,
hadro-dissociation or $p\leftrightarrow n$ conversion becomes more
important than the photo-dissociation.  In addition, if the gravitino
mass is small enough, $\tau_{\tilde{B}}$ becomes shorter than $\sim 1\
{\rm sec}$ and, in such a case, decay of the lightest neutralino
becomes harmless to the BBN scenario.

So far, we have considered the case where the lightest neutralino is
well approximated by the Bino.  We also considered the case where the
lightest neutralino has sizable Higgsino components.  We derived
constraints for the case $U_{\tilde{H}_u}\simeq U_{\tilde{H}_d}\simeq
0.1$, and found that the constraints are almost unchanged compared to
the pure-Bino case.

\subsection{Case with stau NLSP}

Next, we consider the case where the NLSP is charged slepton.  The
basic procedure to derive constraints is explained in
\cite{Kawasaki:2007xb}.  Here, we concentrate on the case where the
NLSP is stau $\tilde{\tau}$, since, in some models of supersymmetry
breaking, like gauge-mediation model \cite{GMSB} and mSUGRA model, 
one of the stau becomes lighter than other sleptons because of the
renormalization group effects.\footnote
{We have checked that the constraints
are almost unchanged when the NLSP is selectron or smuon.}

In considering long-lived stau it is crucial to take account of the
bound state effect which significantly change BBN reaction rates, in
particular, production rate of
$^6$Li~\cite{Pospelov:2006sc,Kohri:2006cn,BSeffects,Cyburt:2006,
Hamaguchi:2007mp,Pradler:2007is,Jedamzik:2007qk}.   In the previous
study by three of the present authors~\cite{Kawasaki:2007xb}, the
number density of $({\rm ^4He}\tilde{\tau}^-)$ bound state was
obtained with the use of Saha equation \cite{Kolb:1990vq}.  After the
study, there have been several works which calculate the number
density of bound state by solving the Boltzmann equation, which gives
more accurate
estimate~\cite{Jedamzik:2007cp,Pradler:2007is,Jedamzik:2007qk}.  In
our current study, we also solve the full Boltzmann equation to
evaluate the number density of bound-state taking effect of
bound-state $({\rm H}\tilde{\tau}^-)$ into account.\footnote
{See also various attempts to tackle this $^{6}$Li problem
\cite{Pradler:2006hh,Buchmuller:2007ui,Bird:2007ge,Jittoh:2007fr,
CBBNsolution,Kusakabe:2007} and a possibility of non-standard $^{9}$Be
production \cite{Pospelov:2007js} in the catalyzed BBN induced by the
stau.}

\begin{figure}
  \begin{center}
    \centerline{{\vbox{\epsfxsize=0.5\textwidth\epsfbox{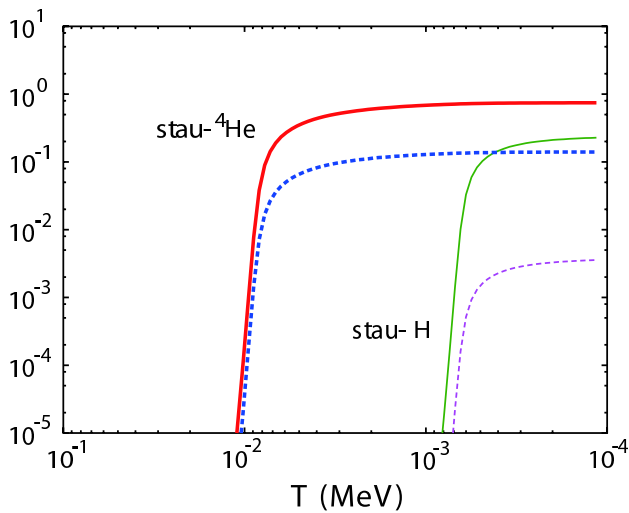}}}}
    \caption{Time evolution of fractions of $\tilde{\tau}^-$ which form
    bound-states $({\rm ^4He}\tilde{\tau}^-)$ (thick solid curve) and
    $({\rm H}\tilde{\tau}^-)$ (thin solid curve). We also show
    $n_4/n_{\rm ^4He}$ (thick dashed curve) and $n_1/n_p$ (thin dashed
    curve). We take $Y_{\tilde{\tau}}= 2\times 10^{-12}$,
    $m_{\tilde{\tau}}= 1$~TeV and $\Gamma_{\tilde{\tau}} = 0$.}
   \label{fig:bound_state}
    \end{center}
\end{figure}

The Boltzmann equations that govern the evolution of the number
densities $n_4$ and $n_1$ of the bound states $({\rm
^4He}\tilde{\tau}^-)$ and $({\rm H}\tilde{\tau}^-)$ are
\begin{eqnarray}
  \frac{d n_4}{dt} & = & - 3Hn_4 -\Gamma_{\tilde{\tau}}n_4
  + \langle \sigma_{r4} v \rangle 
  \left[ (n_{\rm ^4He} -n_4) n_{\tilde{\tau}^{-}} 
   - \left(\frac{m_{\rm ^4He} m_{\tilde{\tau}} T}{2\pi m_4}\right)^{3/2}
   e^{-E_{b4}/T}n_4 \right] , \\
  \frac{d n_1}{dt} & = &  - 3Hn_1 -\Gamma_{\tilde{\tau}}n_1
  + \langle \sigma_{r1} v \rangle 
  \left[ (n_p -n_1) n_{\tilde{\tau}^{-}} 
   - \left(\frac{m_p m_{\tilde{\tau}} T}{2\pi m_1}\right)^{3/2}
   e^{-E_{b1}/T}n_1 \right] ,
\end{eqnarray}
where $n_{\tilde{\tau}^{-}}$ is the number density of the free
negative charged stau, $\Gamma_{\tilde{\tau}}$ is the decay rate of
stau, $E_{b4}\simeq 337.33\ {\rm keV}$ and $E_{b1}\simeq 24.97\ {\rm
keV}$ are the binding energies of $({\rm ^4He}\tilde{\tau}^-)$ and
$({\rm H}\tilde{\tau}^-)$ \cite{Hamaguchi:2007mp}, and $n_{\rm ^4He}$
and $n_p$ are the number densities of $^4$He and proton including both
free and bound states.  Furthermore, $m_4 = m_{\rm
^4He}+m_{\tilde{\tau}}-E_{b4}$ and $m_1 = m_p +
m_{\tilde{\tau}}-E_{b1}$, and the thermally averaged recombination
cross sections for $^4$He and $p$ bound-states are given by
\cite{Kohri:2006cn}
\begin{eqnarray}
   \langle \sigma_{r4} v \rangle & \simeq &
   98.46
   \frac{\alpha E_{b4}}{m_{\rm ^4He}^2\sqrt{m_{\rm ^4He} T}}, \\
   \langle \sigma_{r1} v \rangle & \simeq &
   24.62
   \frac{\alpha E_{b1}}{m_p^2\sqrt{m_pT}} .
%
\end{eqnarray}
An example of the evolution of the abundances of the bound-states
$({\rm ^4He}\tilde{\tau}^-)$ and $({\rm H}\tilde{\tau}^-)$ are shown
in Fig.\ \ref{fig:bound_state}.

\begin{figure}
  \begin{center}
    \centerline{{\vbox{\epsfxsize=0.5\textwidth\epsfbox{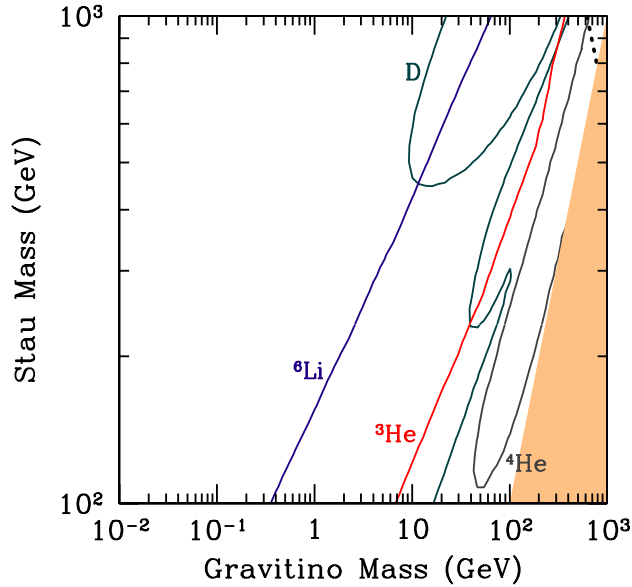}}}}
    \caption{BBN constraints for the stau-NLSP scenario with the
      thermal abundance 
      given in Eq.\ (\ref{Y_stau(thermal)}).}
    \label{fig:stauthermal}
    \end{center}
\end{figure}

First, we use the following thermal relic abundance of $\tilde{\tau}$
\cite{Fujii:2003nr}:
\begin{eqnarray}
  Y_{\tilde{\tau}}
  \simeq 7 \times 10^{-14} \times
  \left( \frac{m_{\tilde{\tau}}}{100\ {\rm GeV}} \right),
  \label{Y_stau(thermal)}
\end{eqnarray}
and derive constraints on $m_{3/2}$ vs.\ $m_{\tilde{\tau}}$ plane.
The result is shown in Fig.\ \ref{fig:stauthermal}.  As one can see,
the most stringent bound is from the overproduction of ${\rm ^6Li}$,
which is due to the catalyzed process.  On the contrary, the bound
from ${\rm D}$ is less severe compared to the Bino-NLSP case.  This is
because, when the stau is the NLSP, the hadronic branching ratio is
several orders of magnitude smaller than that in the Bino NLSP case,
and hence the hadro-dissociation processes are suppressed.

\begin{figure}
  \begin{center}
    \centerline{{\vbox{\epsfxsize=0.5\textwidth\epsfbox{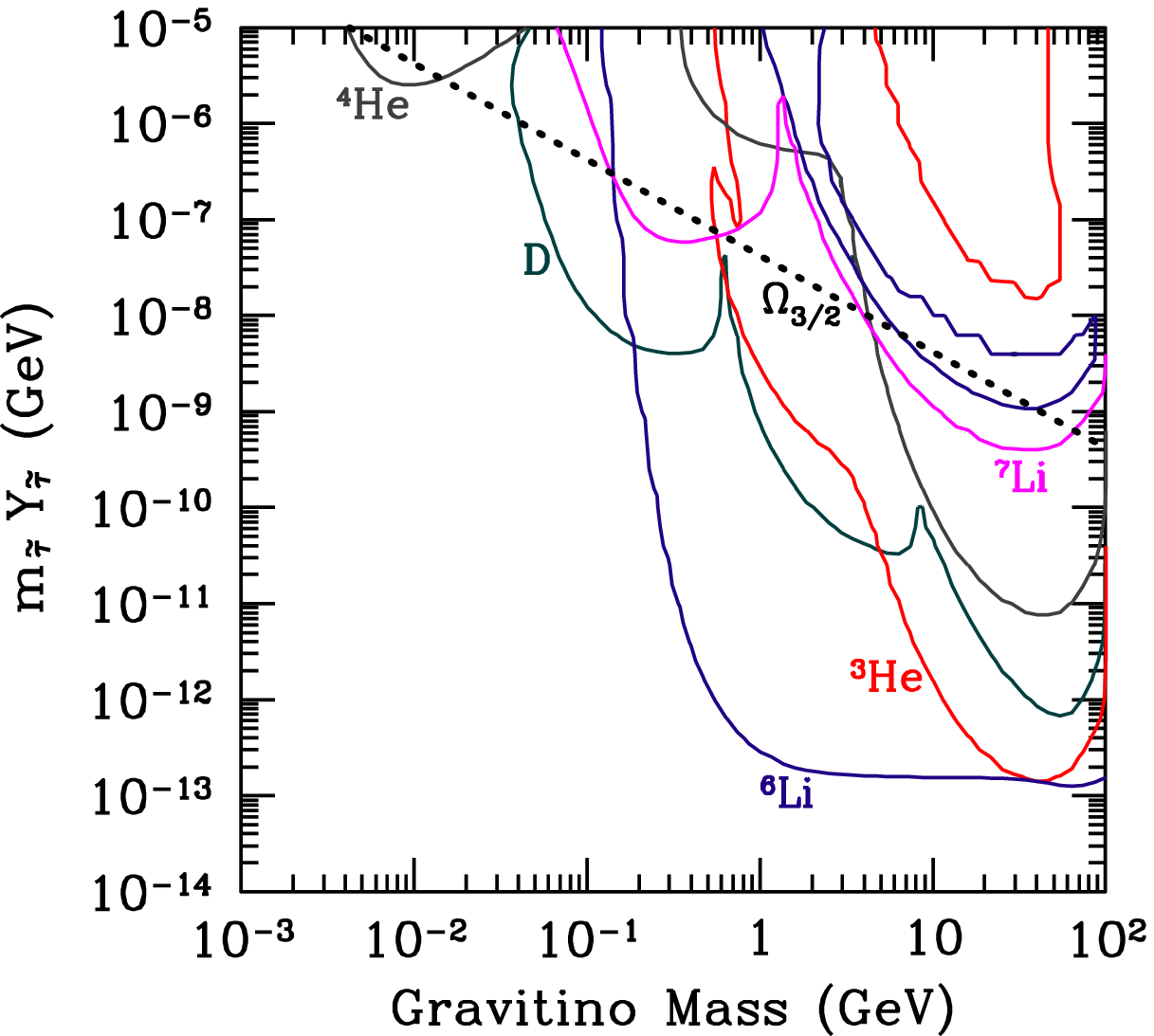}}}}
    \caption{BBN constraints for the stau-NLSP scenario on the
      $m_{3/2}$ vs.\ $m_{\tilde{\tau}}Y_{\tilde{\tau}}$ plane.  The stau
      mass is taken to be $100\ {\rm GeV}$.}
    \label{fig:stauM100}
    \vspace{10mm}
    \centerline{{\vbox{\epsfxsize=0.5\textwidth\epsfbox{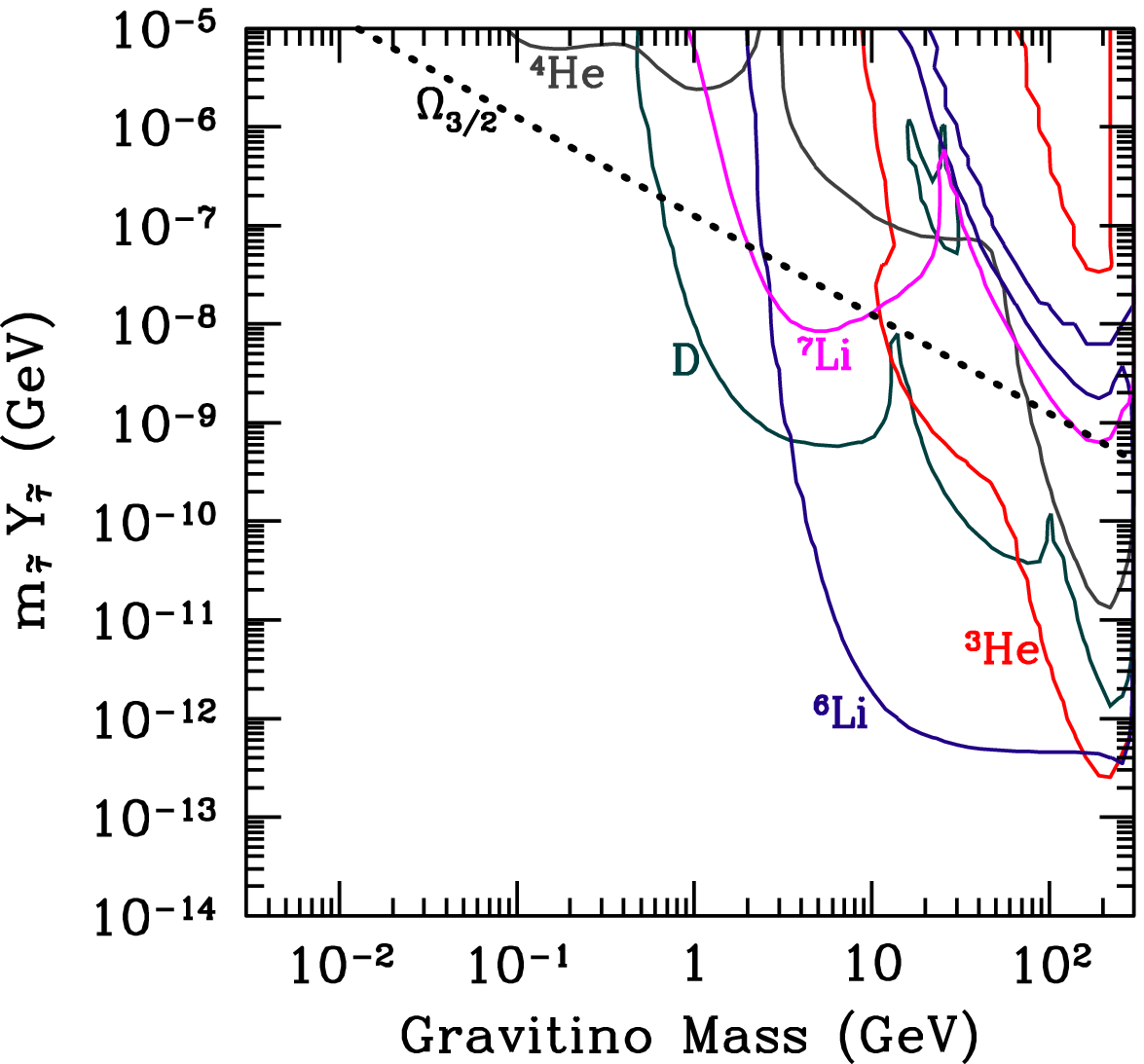}}}}
    \caption{Same as Fig.\ \ref{fig:stauM100}, except for 
      $m_{\tilde{\tau}}=300\ {\rm GeV}$.}
    \label{fig:stauM300}
    \end{center}
\end{figure}

We also treat the primordial abundance as a free parameter and derive
upper bound on $Y_{\tilde{\tau}}$.  The results for
$m_{\tilde{\tau}}=100\ {\rm GeV}$ and $300\ {\rm GeV}$ are shown in
Figs.\ \ref{fig:stauM100} and \ref{fig:stauM300},
respectively.\footnote
{The bound states with singly-charged nuclei such as
(H$\tilde{\tau}^{-})$, (D$\tilde{\tau}^{-})$ or (T$\tilde{\tau}^{-})$
may shield coulomb field of the nuclei completely and significantly
enhance the further reaction rates for these nuclei. These
non-standard processes might totally have reduced the $^{6}$Li,
$^{7}$Li and $^{7}$Be
abundances~\cite{Kohri:2006cn,Jedamzik:2007cp,Jedamzik:2007qk}.
However, the Bohr radius of those bound states are larger than the
typical size of the nuclei (or the square-root of the cross
sections). Then the multi-body problem might be important for
reactions and could not have been understood well. Therefore, we did
not include those effects in the current study.}
The $^{4}$He-constraint for smaller gravitino mass comes from the
$p\leftrightarrow n$ conversion induced by the charged pions which are
mainly produced by the decay of tau lepton \cite{Kawasaki:2007xb}. On
the other hand, the D-constraint comes from the destruction due to
energetic baryons produced by the four-body decay. 
Thus, for the stau case a simple scaling of the constraints on Bino  
by using the hadronic branching ratio does not work.

As we mentioned, in the present study, we have solved the full
Boltzmann equation to calculate the number density of the $({\rm
^4He}\tilde{\tau}^-)$ and $({\rm H}\tilde{\tau}^-)$ bound states.
Then, as discussed in \cite{Pradler:2007is,Jedamzik:2007qk}, the
number density of $({\rm ^4He}\tilde{\tau}^-)$ is reduced compared to
the result with Saha equation, resulting in weaker constraint than the
previous study.  However, the upper bound on $Y_{\tilde{\tau}}$ from
the overproduction of ${\rm ^6Li}$ is increased by the factor $3$ or
so.  Thus, we conclude that the previous study with Saha equation
provided a reasonable estimation of the upper bound on
$Y_{\tilde{\tau}}$.  Notice that, concerning the constraints on the
$m_{3/2}$ vs.\ $m_{\tilde{\tau}}$ plane, the change is
negligible.\footnote
{Ref.\ \cite{Bird:2007ge} suggested a mechanism to reduce the
  $^{7}$Li ($^{7}$Be) abundance through ($^{7}$Be$\tilde{\tau}^-) + p
  \to (^{8}$B$\tilde{\tau}^-) + \gamma$ witch is effective for
  $Y_{\tilde{\tau}} \gtrsim 10^{-11}$ and $\tau_{\tilde{\tau}}
  \lesssim 10^{3}$ sec. Because we have adopted the conservative
  observational $^{7}$Li abundance which agrees with the SBBN
  prediction, this mechanism does not change our results.}

\subsection{Case with sneutrino NLSP}

Finally we study the case with sneutrino LSP. Although such a scenario
is not conventional, it is possible to realize sneutrino MSSM-LSP in
some gravity mediation scenario. In this paper we update the
constraint obtained in \cite{Kanzaki:2006hm} using the most recent
observational data.

\begin{figure}
  \begin{center}
    \centerline{{\vbox{\epsfxsize=0.5\textwidth\epsfbox{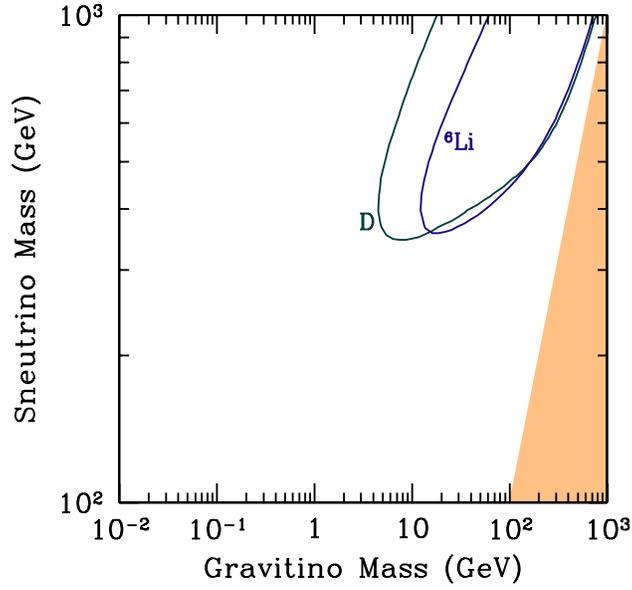}}}}
    \caption{BBN constraints for the sneutrino-NLSP scenario with the
      thermal abundance 
      given in Eq.\ (\ref{Y_snu(thermal)}).}
    \label{fig:snuthermal}
    \end{center}
\end{figure}

The yield variable for thermally produced sneutrino is estimated as
\cite{Fujii:2003nr}
\begin{eqnarray}
    Y_{\tilde{\nu}} 
    \simeq 2 \times 10^{-14} \times
    \left( \frac{m_{\tilde{\nu}}}{100\ {\rm GeV}} \right),
    \label{Y_snu(thermal)}
\end{eqnarray}
where $m_{\tilde{\nu}}$ is the sneutrino mass.  With this yield
variable, we derive the constraint on $m_{3/2}$ vs.\ $m_{\tilde{\nu}}$
plane and the result is shown in Fig.\ \ref{fig:snuthermal}. The
resultant constraint is almost same as that in \cite{Kanzaki:2006hm}
since the most stringent constraints come from overproduction of D and
$^6$Li whose observational abundances have not changed very much.  As
expected, the constraints are much weaker than the Bino- or stau-NLSP
cases because the sneutrino dominantly decays into particles whose
interactions are weak; $\tilde{\nu}\rightarrow\psi_\mu\nu$.  The BBN
constraints are from four-body decay modes
$\tilde{\nu}\rightarrow\psi_\mu\nu f\bar{f}$, which have very small
branching ratios.

\begin{figure}
  \begin{center}
    \centerline{{\vbox{\epsfxsize=0.5\textwidth\epsfbox{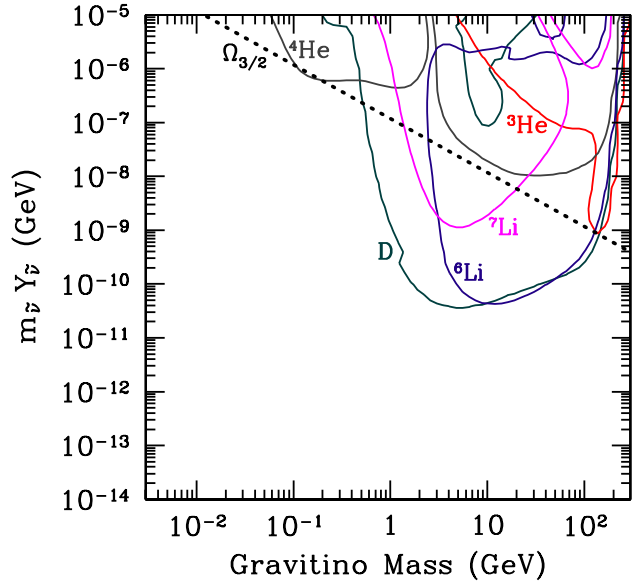}}}}
    \caption{BBN constraints for the snu-NLSP scenario on the
      $m_{3/2}$ vs.\ $m_{\tilde{\nu}}Y_{\tilde{\nu}}$ plane.  The sneutrino
      mass is taken to be $300\ {\rm GeV}$.}
    \label{fig:snuM300}
    \end{center}
\end{figure}

We also treat the primordial abundance as a free parameter and derive
upper bound on $Y_{\tilde{\nu}}$.  The result for
$m_{\tilde{\nu}}=300\ {\rm GeV}$ is shown in Fig.\ \ref{fig:snuM300}.
Compared to the previous study \cite{Kanzaki:2006hm}, it is seen that
the constraint from $^4$He is slightly milder due to the larger
observational constraint on $Y_{\rm p}$ than that adopted in
\cite{Kanzaki:2006hm}.

\section{Conclusions and Discussion}
\label{sec:conclusions}
\setcounter{equation}{0}

In this paper we have obtained the improved BBN constraints on both
unstable and stable gravitino cases.  We have taken into account
recent theoretical and observational progresses in the study of BBN
constraints on long-lived unstable particles.

In the unstable gravitino case, we have included effects of the
longitudinal component which was not considered in the previous
studies, which have led to a more stringent upper-bound on the
reheating temperature for gravitino with relatively small mass.  The
upper bound on the reheating temperature strongly depends on the
gravitino mass and has been summarized in Table \ref{table:bound} for
the MSSM parameters used in our analysis.  We can see that the upper
bound has mild dependence on the mass spectrum of the MSSM particles,
and that the bound becomes almost independent of the MSSM parameters
when the gravitino becomes much heavier than the MSSM superparticles.
Our results give very stringent constraint on the scenario of thermal
leptogenesis \cite{Fukugita:1986hr} which requires the reheating
temperature higher than $O(10^9\ {\rm GeV})$
\cite{Buchmuller:2004nz,Giudice:2003jh}.

When the gravitino is the LSP and stable, the constraints depend on
which superparticle is the NLSP. We have considered Bino, stau and
sneutrino as NLSP and obtained constraints on their properties.  Among
these possibilities, the hadronic branching ratio becomes largest when
the Bino is the NLSP, for which the constraints from the
hadro-dissociation processes are the most stringent.  For the case of
stau as NLSP, we have estimated the effects of bound-state of stau
with $^4$He accurately by solving the Boltzmann equation describing
evolution of the bound-state abundance.  However, the bound from the
catalyzed production process of ${\rm ^6Li}$ is almost unchanged with
the previous study where Saha equation was used, and the
overproduction of ${\rm ^6Li}$ gives stringent bound on the primordial
abundance of stau.  In these two cases, the constraints are so
stringent that the gravitino mass larger than $\sim 10\ {\rm GeV}$ is
excluded when the mass of the MSSM-LSP (i.e., Bino or stau) is lighter
than $1\ {\rm TeV}$.  If the sneutrino is the NLSP, on the contrary,
BBN constraints become drastically weaker since the sneutrino mainly
decays into very weakly interacting particles, i.e., gravitino and
neutrino.  However, in any case, our results provide severe constraint
on some scenario of gravitino dark matter.  Indeed, as we can see in
Figs.\ \ref{fig:binobulk}, \ref{fig:binofocus}, \ref{fig:stauthermal},
and \ref{fig:snuthermal}, the constraints from BBN are severer than
that from the overclosure of the universe.  This fact implies that the
abundance of the gravitino produced by the decay of the MSSM-LSP
cannot be sufficient to realize the gravitino dark matter if the
thermal relic abundance of the MSSM-LSP is assumed.  Thus, in order to
realize the gravitino dark matter, most of the gravitino should be
produced by, for example, the scattering processes of thermal
particles at the time of the reheating after inflation
\cite{Moroi:1993mb} or by the decay of scalar condensation
\cite{GravFromModuli,GravFromInflaton}.

\vskip 0.5cm

\noindent 
{\it Acknowledgements}: This work was supported in part by PPARC grant
PP/D000394/1, EU grants MRTN-CT-2004-503369 and MRTN-CT-2006-035863, and
the European Union through the Marie Curie Research and Training Network
``UniverseNet'' (K.K.).  This work was also supported in part by the
Grant-in-Aid for Scientific Research from the Ministry of Education,
Science, Sports, and Culture of Japan, No 14102004 (M.K.) and No.\
19540255 (T.M.), and JSPS-AF Japan-Finland Bilateral Core Program
(M.K.).


\end{document}